# On the Role of Charge Transfer Excitations in Non-Fullerene Acceptors for Organic Photovoltaics

Samuele Giannini,*[a,g] Daniel J. C. Sowood,[b] Jesús Cerdá,[a] Siebe Frederix,[c] Jeannine Grüne,[d] Giacomo Londi,[a,h] Thomas Marsh,[b] Pratyush Ghosh,[b] Ivan Duchemin,[f] Neil C. Greenham,[b] Koen Vandewal,[c] Gabriele D'Avino,[e] Alexander J. Gillett,*[b] David Beljonne*[a]

a) Laboratory for Chemistry of Novel Materials, University of Mons, Mons 7000, Belgium.
b) Cavendish Laboratory, University of Cambridge, JJ Thomson Avenue, Cambridge, CB3 0HE, U.K.
c) Instituut voor Materiaalonderzoek (IMO-IMOMEC), Hasselt University, Diepenbeek, Belgium
d) Julius Maximilian University of Würzburg Am Hubland, Würzburg 97074, Würzburg, Germany
e) Grenoble Alpes University, CNRS, Grenoble INP, Institut Néel, 38042 Grenoble, France
f) Grenoble Alpes University, CEA, IRIG-MEM-L_Sim, 38054 Grenoble, France
g) Present address: Institute of Chemistry of OrganoMetallic Compounds, National Research Council (ICCOM-CNR), I-56124 Pisa, Italy
h) Present address: Dipartimento di Chimica e Chimica Industriale, Universitá di Pisa, 56124 Pisa, Italy.

e-mail: samuele.giannini@umons.ac.be, ajg216@cam.ac.uk, david.beljonne@umons.ac.be

## Highlights

- A detailed analysis of the steady-state and transient optical absorption spectra of the non-fullerene acceptor Y6 thin films reveals clear evidence for mixing between localized Frenkel and intermolecular charge-transfer excitations resulting in a broad density of states.
- These hybrid states are in dynamic equilibrium with bound polaron pairs that predominantly recombine on timescales much shorter than that for charge separation in blends of Y6 with polymer donors.
- Our combined experimental-theoretical study clarifies the origin for the limited charge generation yield in single component Y6 organic solar cells and highlights the importance of molecular packing in defining the nature of the primary photoexcitations in non-fullerene molecular acceptors.

## Abstract

Through the development of new non-fullerene electron acceptor (NFA) materials, such as Y6 and its molecular derivatives, the power conversion efficiencies of organic photovoltaics (OPVs) have now exceeded 19%. However, despite this rapid progress, our fundamental understanding of the unique optical and electronic properties of these Y-series NFAs is lacking, and this currently limits progress in material design. In this work, we provide a detailed computational-experimental characterisation of the archetypal NFA, Y6. To explain the significant broadening and red shift of the absorption spectrum observed when moving from the solution phase to the solid state, we first rule out more typical causes, such as J-aggregation. Instead, by considering the role of charge transfer (CT) excitations and their mixing with Frenkel exciton (FE) states, we can computationally reproduce the experimental absorption spectra of Y6 with excellent accuracy. Using transient absorption spectroscopy, we provide evidence for this dense manifold of FE-CT hybrid electronic excitations in Y6 through the prominent sub-picosecond relaxation events following supra band gap excitation. Furthermore, through sub band gap excitation, we also find states with polaronic character in Y6 that are in a dynamic equilibrium with the FE-CT hybrid states. Magnetic resonance spectroscopies reveal that these polaronic states are polaron pairs, most likely located on neighbouring Y6 molecules, not free charge carriers, as has been previously suggested. Thus, this new understanding of how the solid-state packing motif directly controls the optical and electronic properties of Y-series NFAs opens the door to intelligently design NFA materials to further increase OPV performance.

## Introduction

Significant progress has been achieved in the field of organic photovoltaics (OPVs) due to extensive synthetic work, and this has now resulted in record-breaking power conversion efficiencies exceeding 19%.[1,2] A particular family of molecular non-fullerene acceptors (NFAs), *i.e.*, Y6 and its analogues, has dominated the field since 2019[3,4] and has been central to the recent efficiency developments.[5,6] However, despite the major performance advances enabled by Y-series NFAs, the fundamental photophysics of Y6 and its molecular derivatives is still not well understood and there is considerable debate in several areas of great relevance for OPV operation.[7–10] For example, there is no consensus yet on the reason for the significant red shift and broadening of the Y6 absorption spectrum when moving from solution to thin-film,[8,10] nor on whether Y6 can spontaneously generate free charge carriers in the bulk acceptor phase.[7,9,11,12] Therefore, gaining a fundamental understanding of the optical properties and establishing structure-photophysical property relationships would provide valuable insights for knowledge-driven progress, ultimately pushing the efficiency of these materials beyond the 20% milestone.

What makes the Y-series NFAs so special? Unlike previous generations of NFAs, materials belonging to the Y-series are based on a general A–DA'D–A structure (A = acceptor moiety, D = donor moiety). The A' unit leads to a centrosymmetry loss relative to the previous A–D–A NFAs and the formation of a permanent dipole moment that encourages dimer formation in the crystal structure. However, the performance of OPVs is not solely determined by the chemical building blocks themselves but is also highly sensitive to the arrangement and orientation of the individual molecules in the solid-state.[13] For instance, it has been shown that, compared to earlier NFA molecular families (such as ITIC or IEICO), Y6 assembles into a unique 3D interpenetrated lattice. This structure, of which variations are common to other Y-series derivatives, shows a large number of directional face-on stackings yielding an interconnected transport network with sizable charge and excitonic interactions.[14] Importantly, this characteristic molecular crystal packing is essentially retained in thin-film phases after processing.[15–17] This local microstructure dictates the nature of the relevant photo-excitations in both the bulk phase of the acceptor component and at the heterojunction with a donor component, with a possible mixing between localized Frenkel exciton (FE) and charge transfer (CT) excitations and a degree of wavefunction delocalization that is limited by (static and dynamic) energetic disorder.[18,19] These characteristics in turn define the optical properties and excited-state dynamics in these systems. For instance, a remarkable feature of Y6 is the dramatic change in position and shape of its spectral features upon aggregation. Specifically, worth noting is the unusually strong spectral red-shift (by ~0.2 eV) and increased broadening when going from Y6 solution to the thin-film phase.

Curiously, recent experimental work has shown that neat Y6 can work reasonably well as a single component solar cell, with a power conversion efficiency of ~4.5%.[11] However, open questions still remain about whether Y6 crystalline domains are intrinsically generating a significant quantity of free charge carriers, or whether charge separation is happening at the domain boundaries of these 'single component' devices (potentially assisted by the electric field applied across the OPV device under operation'),[20] with the device in-effect operating as a bilayer OPV. Spectroscopic and theoretical investigations on this topic point to the presence of closely interacting molecules that can yield the formation of charge transfer/separated states as precursors for possible free carrier generation.[7–10] Wang et al. were the first to infer the existence of a delocalised excited state, which they called an 'intra-moiety state', from transient absorption spectroscopy (TAS) measurements on neat Y6 films as well as PM6:Y6 bulk heterojunction (BHJ) blends.[9] With selective photoexcitation of Y6 in the BHJ blend, they showed that this state is an intermediate on the pathway of hole transfer to the polymer. While the exact nature of the delocalised state was not explicitly defined, it exhibited characteristics suggestive of electron-hole separation resembling that of a polaron-pair-type state. They also assigned a photo-induced absorption (PIA) feature in their TAS spectrum of a neat Y6 film at 1550 nm to this state, which was absent in solution measurements of Y6. Building on Wang et al.'s work, Price et al. made an eye-catching claim about the spontaneous generation of free charges with a 60-90% yield in neat Y6 films.[7] This observation was based primarily on intensity-dependent external photoluminescence quantum efficiency (PLQE) measurements, where an increase in PLQE with fluence was attributed to radiative, non-geminate recombination of intrinsically generated free charges. However, it is difficult to reconcile the formation of truly 'free' charge carriers (*i.e.*, exceeding the Coulomb capture radius, typically of ~5 nm in organic semiconductors[21]), with the widely reported excited state lifetime of a few hundred picoseconds in neat Y6 films, as measured by TAS.[9] Though, it is also worth noting that transient PL studies point to a longer fluorescence lifetime on the order of one nanosecond in neat Y6 films.[22,23] Thus, there is clearly significant confusion surrounding the excited state dynamics of Y6 and related molecular derivatives that needs to be addressed.

In this work, we investigate the role played by CT excitations within the bulk NFA phase in the photophysics of Y6 (Fig. 1a,b). Using state-of-the-art modelling combining high-quality electronic structure calculations,[24–26] polarizable molecular embedding[27–29] and a Frenkel-Holstein-like exciton Hamiltonian,[30–32] we show that the optical absorption of Y6 films is indeed largely perturbed by intermolecular interactions. We also computationally address the nature of the electronic excitations in neat Y6 films, specifically on the role and interplay between FE and CT excitations. Using TAS and magnetic resonance spectroscopy, we corroborate the findings of the quantum-chemical calculations and investigate the potential for polaronic states to form in neat Y6 films. We reveal that these polaron pair states are most probably located on neighbouring Y6 molecules and proposed that they are not free carriers as previously suggested. Thus, by providing an understanding of the intermolecular factors that are responsible for the unique optical and electronic properties of aggregated Y6, our work sets out a blueprint for intelligently designing new Y-series NFAs for OPV applications. We believe that this can be achieved by using chemical modifications to the Y-series molecular backbone to engineer the critical solid-state packing motif which can, in turn, be exploited to control the properties in the bulk NFA phase.

## Results and Discussion

### Y6 steady-state experimental absorption.

We start our analysis by investigating salient features characterizing the absorption spectra of Y6 in dilute solutions of Toluene (TOL) and Chloroform (CF). We report the measured solution spectra in Fig. 1c and give details about the experimental measurements in the Methods. In both solvents, we observe an intense absorption peak corresponding to the transition energy, $E_{S_0 \to S_1}^{(sol)}$, from the ground ($S_0$) to the first excited state ($S_1$) and the associated vibronic progressions. $E_{S_0 \to S_1}^{(sol)}$ is 712.5 nm (1.74 eV) in TOL and 733.6 nm (1.69 eV) in CF (purely electronic, 0-0 transition). The additional band ~0.40 eV higher in energy than $S_0 \to S_1$ is related to a second electronic transition (*i.e.*, $S_0 \to S_2$) characteristic of the $C_{2v}$ symmetry point group of the Y6 molecule, as observed in previous work.[33,34] This is confirmed, as shown in the Supplementary Information (SI) Section S3, by employing a Franck-Condon (FC) fitting model expanded to include multiple electronic transitions and vibrational modes for each electronic excitation.[33] Interestingly, in Fig. 1c, we further observe a modest positive solvatochromism upon increasing the polarity of the solvent (*i.e.*, a red-shift going from TOL with a dielectric constant of 2.38 to CF with a larger dielectric constant of 4.81), which is typical of push-pull dyes characterized by a small degree of intramolecular charge separation in the ground state.[35] Associated to such a red-shift we can also see a smearing out of the vibronic spectral features, though the ratio between the intensity of the first vibronic 0-1 band and the main transition, the so-called Huang-Rhys factor ($S_{eff} = \frac{I_{1-0}}{I_{0-0}}$)[31], is not significantly affected (see Table 1).

We now turn to the analysis of the experimental absorption spectrum of Y6 thin-film phase prepared by spin casting the material from solution on the substrate, as described in the Methods. A number of works[16,17] have shown that a Y6 film processed from CF solution maintains an extended crystalline structure and adopts a face-on orientation on the substrate in its thin-film phase, thereby making such a system an ideal test for our simulations which are performed on 'perfect' single crystals (see Section S1).[36] The CF-processed Y6 thin-film absorption spectrum is reported in Fig. 1c. The transfer matrix modelling (TMM) approach developed by Kerremans et al.[37] has been used here to correct the absorption coefficient for interference and thickness dependence effects (see Methods and Section S12 for details). By comparing the thin-film spectrum with the corresponding Y6 spectrum in CF solution (Fig. 1c), we clearly see three noticeable features: (i) a remarkably large red-shift (~0.18-0.23 eV) upon aggregation, (ii) an increase of the 0-1 band by more than a factor of two in the thin-film phase with respect to the solvated single molecule and (iii) a significant broadening of the Y6 aggregate absorption spectrum. The red-shift going from solution to the thin-film phase is a common feature of ideal J-type excitonic interactions where the optically allowed state is at the bottom of the excitonic band of the aggregate.[32] The width of the related energy band depends on the strength of the excitonic interactions between the molecules. For ideal J-aggregates, the stronger these interactions, the more red-shifted the spectrum. At first, it may seem plausible to assign the strong red-shift (feature (i)) to strong J-like interactions in Y6 aggregates. However, ideal J-aggregates actually show a decrease in the intensity of the vibronic 0-1 side band (relative to 0-0) and a reduction in spectral broadening.[32] Both these characteristics are in contrast with features (ii) and (iii) observed in Y6 aggregates. Additionally, a number of experimental investigations[7,9,11,38] anticipate that potential Y6 intermolecular CT states might interact with Frenkel-type excitonic states. This could have an impact on the optical properties and the shape of the spectra, as previously observed in other systems.[31,39–41] To understand all these peculiarities of Y6 steady-state optical properties, we turn to state-of-the-art excited-state electronic structure calculations connected to large-scale model Hamiltonians that are necessary to describe extended aggregates.

**Y6 steady-state simulated absorption.**

We have simulated the spectral features of Y6 in solution using high-quality Bethe Salpeter equation (BSE) calculations with the inclusion of a polarizable continuum model (PCM) to simulate the effect of the solvent, as detailed in Methods.[24,25] In Fig 1d, we achieved a remarkably good agreement (within ~0.1 eV tolerance) with respect to measured absolute excitation energies in both TOL and CF solutions. BSE/PCM provides accurate vertical excitation energies ($E'^{,(\text{sol})}_{S_0 \rightarrow S_1}$) as reported in Table 1 and is able to accurately recover the solvatochromic red-shift observed in the experiment by changing the dielectric constant of the medium. From the vertical excitation energies in both solvents, the corresponding adiabatic energies, to which the main peak ($E^{(\text{sol})}_{S_0 \rightarrow S_1}$) of the simulated spectra is referred, can be estimated by subtracting the relaxation energy of the $S_1$ state from $E'^{,(\text{sol})}_{S_0 \rightarrow S_1}$. The relaxation energy is evaluated in the harmonic approximation as explained in Section S3 and it is decomposed in terms of frequency-resolved normal modes (see Fig. S3). As done in previous works,[31] we assume that such a relaxation is driven by a single effective vibrational mode that couples the excitation to $S_1$ state (Fig. 1d) and is responsible for the vibronic progression observed in Fig. 1c. For this effective mode we computed a frequency, $\hbar\omega_{\text{eff}} = 0.18$ eV, characteristic of double C=C bond stretching/aromatic ring breathing (common to other π-conjugated organic systems[31]). The calculated value agrees with resonance Raman experiments that reveal two prominent peaks in the spectrum of Y6 at 0.18 eV and 0.19 eV, respectively (see Fig. S12). This analysis yields an effective Huang-Rhys factor ($S_{\text{eff}}$) associated to $S_0 \rightarrow S_1$ excitation of about 0.33, in line with the experimental observation (see Table 1). By considering the calculated $\hbar\omega_{\text{eff}}$ and $S_{\text{eff}}$ and using Eq. S5, we reproduced the vibronic progression of the main $S_0 \rightarrow S_1$ transition in Fig. 1d. We found a good qualitative agreement between the spectral shape of the simulated solution spectra in both solvents with respect to experiments, and we remark that no additional shift has been applied to our simulated spectra. Although in principle we could systematically improve the agreement between the vibronic shape of the simulated solution spectrum, for instance including multiple vibrational modes and electronic transitions (see Fig. S4b), our ultimate goal is to simulate the spectrum of an extended thin-film system. To this end and to keep the model computationally tractable when approaching large systems, considering only a single effective mode per molecular site is a widely used and sensible approximation for this system.[31,32]

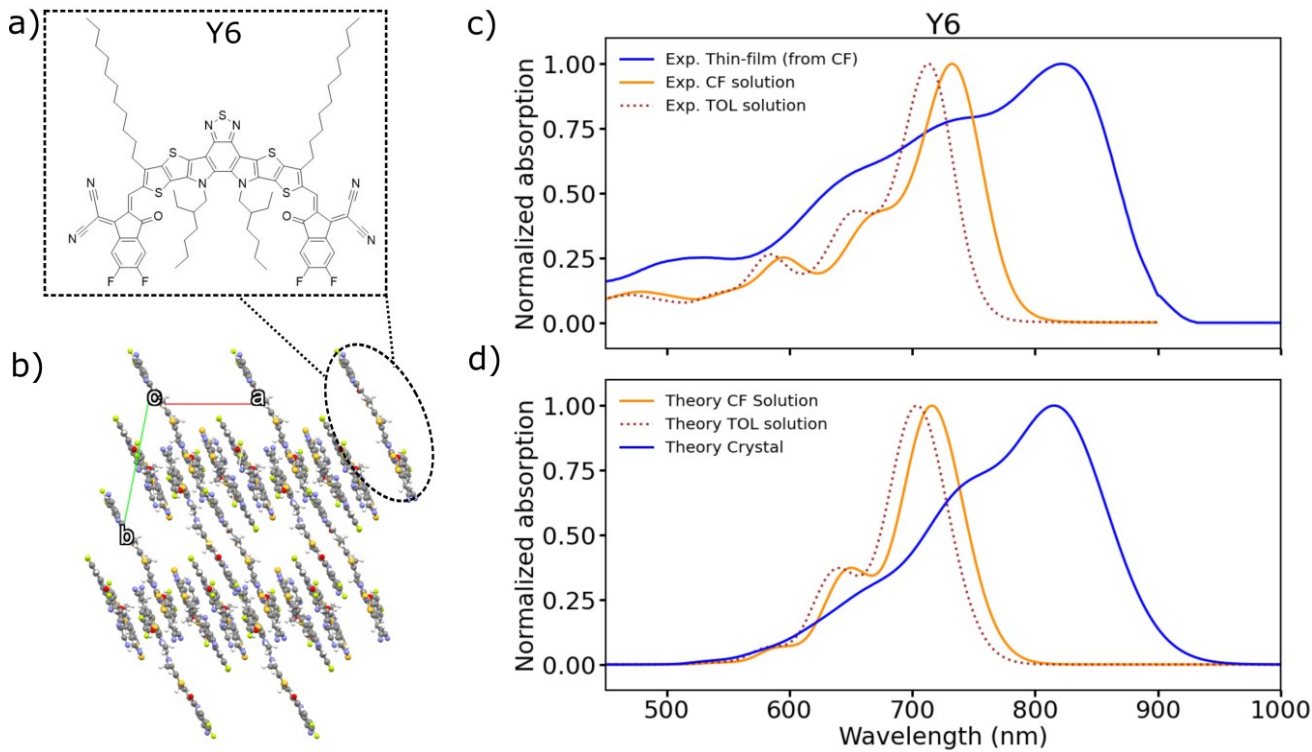

**Figure 1: a)** Molecular and **b)** single-crystal structures of Y6, respectively. The alkyl side chains have been removed for clarity in the crystal structure representation. The unit cell axes a and b are shown in red and lime, respectively (axis c is eclipsed by the other two). **c)** Experimental absorption spectra of Y6 in TOL (dashed brown line) and CF solution (solid orange line). CF-processed thin-film spectra obtained applying the correction for interference effects using the transfer matrix method (TMM) as described in the text.[37] **b)** Computed normalized absorption spectra of Y6 measured in TOL (dashed brown line) and CF (solid orange line). The energies of the $S_0 \rightarrow S_1$ were obtained from BSE/PCM calculations and the vibronic progressions calculated as described in Methods. The spectrum of Y6 crystal is reported with solid blue line obtained using the Frenkel-Holstein excitonic Hamiltonian (Eq. 1) parameterized using ab-initio data as described in the text. A homogeneous spectral broadening of 60 meV was added to all computed spectra for the best comparison with experiments.

**Table 1:** Comparison between measured and computed experimental features of Y6.

| Experiments | Toluene | CF |
|---|---|---|
| $E_{S_0 \rightarrow S_1}^{(sol)}$ [eV] | 1.74 | 1.69 |
| $E_{S_0 \rightarrow S_1}^{(thin-film)}$ [eV] | 1.51 | |
| Total Red-shift [eV] | -0.23 | -0.18 |
| $S_{eff}$ solution | 0.37 | 0.37 |
| **Theory** | **Toluene** | **CF** |
| $E_{S_0 \rightarrow S_1}^{\prime,(sol)}$ [eV][a] | 1.82 | 1.79 |
| $E_{S_0 \rightarrow S_1}^{(sol)}$ [eV][b] | 1.76 | 1.73 |
| $\Delta_{0-0}$ [eV][c] | -0.19 | -0.16 |
| $E_{FE}^{(cry)}$ [eV][d] | 1.57 | |
| Aggregation Shift [eV][e] | -0.050 | -0.050 |

| $E_{S_0 \to S_1}^{(cry)}$ [eV] | 1.52 | |
|---|---|---|
| Total Red-shift [eV] | -0.24 | -0.21 |
| $S_{eff}$ solution | 0.33 | |

[a] Vertical excitation energy in solution. [b] Adiabatic excitation energy obtained subtracting the relaxation energy ($\lambda_{hf}^{rel}$ in Table S1) evaluated in the harmonic approximation from the vertical excitation energy as described in the text. [c] Solution-to-Crystal shift (STCS) evaluated as $\Delta_{0-0} = E_{FE}^{(cry)} - E_{S_0 \to S_1}^{(sol,CF)}$ as described in the text and Methods. [d] Site energy relative to the excitation of a single molecule embedded in the crystal (see Methods and Section S8). [e] Red-shift due to aggregation effects and mixing between FE and CT states. The minus signs appearing in the energy values indicate a shift toward lower energy.

Compared to the solution case, the description of the optical properties and the electronic excitations in solid state Y6 (Fig. 1b) is complicated by the presence of intermolecular interactions and the larger dimensionality of the system. To accurately describe the spectrum in the crystalline form and explain the spectral features observed in experiments, we used a powerful Frenkel-Holstein Hamiltonian[30–32] (Eq. 1) supplemented by *ab initio* calculations, as described in the Methods section. In brief, such a Hamiltonian is represented on a diabatic basis of localized electronic states of different characters, *e.g.*, single FE states, where electron and hole are on the same site, CT excitons, where electron and hole sit on different sites (Fig. S5), and it explicitly includes the electronic interactions between them. Namely, the long-range excitonic $V_{kl}$ couplings between tightly bound electron-hole pair sitting either on molecules $k$ or $l$, which determine exciton delocalization; photoinduced electron transfer couplings ($D_e$) and hole transfer couplings ($D_h$) that allow for exciton splitting into CT excitons; and electron and hole transfer integrals ($t_e$ and $t_h$, respectively), which describe the hopping of charges between neighboring molecules, possibly leading to charge separation (see representation in Fig. S5). This Hamiltonian also incorporates the Coulomb attraction between electron and hole as a function of distance, as well as the coupling between electronic and nuclear degrees of freedom. This vibronic coupling interaction is introduced in the model by considering effective modes coupled to the creation of excitonic, cationic and anionic states, as described in Section S3 and Methods.

Holstein-like Hamiltonians such as the one used in this work have been employed with great success Holstein-like Hamiltonians as the one used in this work have been employed with great success in previous work[30–32] to describe the optical properties of aggregates and crystals of conjugated organic molecules,[32,41,42] supramolecular systems,[43] and extended polymers[44,45] and copolymers[46]. However, the novelties of our computational framework over previous work are several, the most important being the following. (i) We move away from ideal lattices of reduced dimensionality with only nearest neighbour interactions,[32] in favour of the three-dimensional morphology determined by X-ray diffraction.[36] (ii) The excitation energies that constitute an important part of the Hamiltonian are computed using state-of-the-art BSE calculations coupled to classical polarizable embedding models of atomistic resolution (MM). This formalism allows us not only to retrieve the energy of the FE states, as discussed for the single molecule in solution, but also to accurately determine the energetic position of CT excitations that are notoriously challenging to capture with conventional time-dependent density functional theory (TDDFT).[47] Furthermore, this approach also permits to accurately account for environmental (electrostatic and dielectric screening) effects,[24,25,27–29] which are of great importance for our analysis, for instance, in the remarkable solution-to-crystal shift described below.

(iii) We account for the complete range spanned by the long-range excitonic interactions (*i.e.*, beyond nearest neighbour interactions); all being computed from first principles methods that account for the atomistic details of the interacting molecules. The proposed theoretical protocol, detailed in Methods, represents an important step towards a realistic description of the electronic states and optical properties of application-relevant molecular extended systems.

Once constructed, the Hamiltonian is diagonalized to determine the excitonic states needed to generate the absorption spectrum of the Y6 single crystal (Eq. S4). The Y6 crystal structure (Fig. 1b) used for the calculations is described in Section S1. The resulting theoretical spectrum is presented in Fig. 1d for a 3x3x3 supercell in periodic boundary conditions and compared to the experimental counterpart in Fig. 1c. We note that the agreement both in terms of peaks position and relative intensity is excellent. We were able to capture all the key features observed in experiments, including the significant red-shift (calculated to be 0.21-0.24 eV) between the CF or Toluene solutions and solid-state phase, and the corresponding enhancement of the 0-1 vibronic band. These characteristics are quantitatively reproduced by our simulations (see Table 1). Furthermore, we successfully reproduced the larger broadening of the solid-state spectrum compared to the solution counterpart. Convergence of the results in terms of system sizes, robustness with respect to electronic couplings and other technical aspects are shown in Fig. S9. Moving forward, we will discuss the influence of various components in the Hamiltonian and analyse their contribution to the overall shape of the spectrum to better understand the electronic states of the Y6 thin-film phase.

**Aggregation effect in Y6: H-like vs J-like interactions.**
We start our detailed theoretical analysis focussing on the Frenkel exciton block ($H_{\text{FE}}$ in Eq. 2) of the Hamiltonian to investigate whether the red-shift from solution to the aggregate is due to Kasha's aggregation effects and strong J-like excitonic interactions. To this end, we computed the off-diagonal elements, $V_{kl}$, between pairs of molecules $k$ and $l$ extracted from a supercell of Y6. $V_{kl}$ excitonic interactions are approximated using the long-range Coulomb component between transition densities,[48] which scale as $r^{-3}$ ($r$ is the intermolecular distance).[18,48,49] In this work, we used transition electrostatic potential (TrESP) charges to represent molecular transition densities (Eq. S6). Notably, this approach allows to go beyond the commonly used point-dipole approximation which is known to result in large overestimation of the excitonic couplings in closely packed solids of large molecules, like Y6.[18] Parametrization details of TrESP charges and benchmarks on the accuracy of this approach compared to robust multi-state fragment-excitation charge difference are given in Section S6. The excitonic coupling values for the closest contact nearest neighbour pairs are reported in Table S2; they are all sizable, pointing to a strong excitonic effect in Y6. Besides the nearest neighbour, our fast TrESP approach allows us to compute the full range of excitonic interactions in large Y6 crystals quickly and accurately. The use of atomic TrESP charges is also useful to ensure a consistent relative sign of the interaction within the Hamiltonian in Eq. 2. We discuss this important point in more details in Section S6. In Figs. 3a,b, we visually show the amount and strength of J-like interactions in blue vs H-like interactions in red on a supercell. Notably, H-like interactions are abundant in Y6 crystals, about 64% for this 3x3x2 cell size, compared to 36% of J-like interactions. This hints to the fact that the red-shift observed in Fig. 1c might not be a consequence of J-aggregation.

The excitonic eigenstates of $H_{\text{FE}}$ were used to calculate the Frenkel exciton spectrum of the crystal according to Eq. S4. The excitonic couplings refer to the crystal structure using periodic boundary conditions with a minimum image convention to deal with extended systems. These interactions were screened by an isotropic dielectric constant ($\varepsilon$ = 2.9), which we computed with the same MM model employed in embedded *ab initio* calculations.[29] This value is consistent with what was measured by Li et al. in Ref.[50] ($\varepsilon$ = 3.36). A detailed description of these calculations is provided in Methods. Besides excitonic couplings, we also accurately computed the solution-to-crystal shift (STCS), $\Delta_{0-0} = E_{\text{FE}}^{(\text{cry})} - E_{S_0 \to S_1}^{(\text{sol,CF})} = -0.16$ eV when going from CF solution to the crystalline phase. Both the solution and the crystal excitation energies has been obtained with embedded BSE calculations, the former adopting a PCM embedding mimicking CF, the second describing the crystalline environment with an atomistic polarizable MM embedding constituting the crystalline environment. The significant STCS mostly arises from the higher dielectric constant of the solid, with respect to the CF solution, plus smaller contributions from electrostatic crystal fields and molecular geometries (see Table S6). Computed energy values are reported in Table 1 and the protocol used for these high-quality calculations is described in Methods.

The FE spectra of the Y6 aggregate without and with vibronic coupling are reported in Figs. 2c and 2d, respectively, and compared to the solution spectrum in CF. We can observe that in both cases the Y6 crystal is red-shifted compared to solution by an amount that essentially corresponds to the STCS ($\Delta_{0-0}$). The similar amount and strength of J-like vs H-like interactions makes it such that the effect of aggregation in Y6 is almost null with only a tiny (3 meV) shift towards the red edge of the spectrum compared to the vertical black dashed line in Fig. 2c (representing $E_{\text{FE}}^{(\text{cry})}$). Fig. 2d shows that considering the vibronic coupling to FE states results in a slight improvement of the overall spectral shape. The coupling to an effective vibration yields the formation of the 0-1 side band observed in experiments. However, the intensity of this vibronic replica is much lower than the 0-0 band, while comparable intensities have been measured in the experimental absorption spectrum of Y6 films (Fig. 1c). Moreover, although the STCS is substantial, the total red-shift going from the solution phase to the aggregate is lower than what we measured. These observations suggest that other contributions, in addition to aggregation and coupling to vibrations of bright FE are still missing from this preliminary analysis.

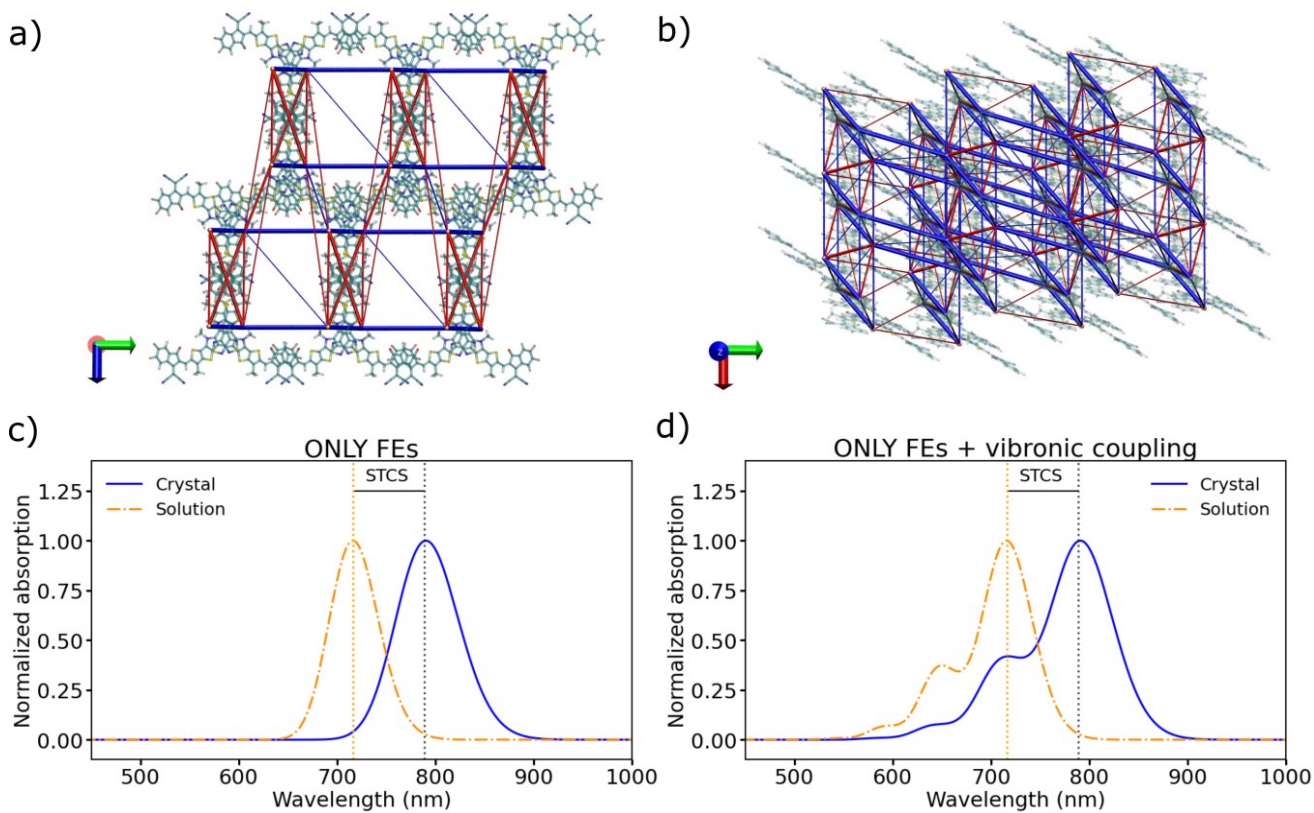

**Figure 2: (a)** Top and **(b)** lateral view of Y6 crystals where H-like (J-like) interactions are shown as blue (red) segments. Only interactions stronger than 15 meV are shown for clarity purpose and the thickness of the coloured lines is proportional to the excitonic coupling strength. In this picture, the H-like (J-like) interactions are defined according to the positive (negative) sign of the transition-dipole-corrected coupling $\tilde{V}_{kl} = \text{sgn}(\boldsymbol{d}_k \cdot \boldsymbol{d}_l)V_{kl}$, where the term in between parenthesis is the scalar product of the transition dipoles obtained using atomic TrESP charges $\boldsymbol{d} = \sum_I q_I^T \boldsymbol{r}_I$ ($I$ runs over the atomic positions of the molecule and $q_I^T$ are the atomic TrESP charges). Panels **(c)** and **(d)** show the solution and crystal absorption spectra computed with a pure Frenkel exciton model ($H_{\text{FE}}$ in Eq. 2) with and without vibronic coupling, respectively. The solution to crystal shift (STCS) is indicated in black. Dashed black vertical lines represent the FE energies ($E_{\text{FE}}^{(\text{cry})}$) of a molecule in the solid, while the dashed orange ones represent the excitation energies in CF solution ($E_{S_0 \to S_1}^{(\text{sol,CF})}$). Both values are obtained with BSE as explained in the text. No shift is applied to the computed spectra.

**Intermolecular CT excitations: impact on the optical properties of Y6.**
The nature of the primary electronic excitations in thin-films might change considerably upon switching on the interactions between FE and CT states and this should have an impact on the spectral shape. Recent works showed that this might be the case in pristine Y6,[7,10] though the lack of vibronic effects, and the use of routine TD-DFT calculations, whose results are notably functional dependent when it comes to CT exciton energies, prevented from reaching reliable conclusions. We thus completed the Hamiltonian in Eq. 1 calculating $H_{\text{CT}}$ and $H_{\text{FE-CT}}$ (Eqs 3 and 4) matrix elements to account for the formation and mixing between CT and FE states. To this end, we evaluated photoinduced electron and hole couplings and hole and electron transfer integrals ($D_e$, $D_h$, $t_e$ and $t_h$, respectively) between the molecules in the crystal (see Fig. S5) and used accurate diabatization approaches as described in Sections S6 and S7.

Because all these interactions decrease exponentially in magnitude with increasing distance, as they essentially depend on the orbital overlap of the interacting molecules, we computed them in Y6 for all the pairs within a maximum distance cut-off of 20 Å from each other. Among all the pairs within the distance threshold depicted in Fig. S2, six pairs with a minimum interatomic distance below 3.5 Å exhibit non-negligible (*i.e.*, above 0.1 meV) electronic couplings (see Table S3). It is worth noticing that these interactions are in some cases larger in magnitude than related excitonic couplings, pointing to strong possible mixing between FE and CT excitations. Nonetheless, it is also crucial to quantify the energy of the CT states to establish to which extent the energy level alignment compared to the couplings between FE and CT states is sufficiently favourable to modify the spectral shape and photophysics of the crystal.[40] To this end, we quantified the energy of CT-like excitations as a function of electron-hole distance, $E_{CT}(r_s)$, using BSE calculations with polarizable MM embedding performed on some of the representative pairs extracted from the crystal. From these data, we derived a (screened) Coulomb barrier (see Fig. S8), describing the energetics of electron and hole particles at various distances and completed the Hamiltonian in Eq 1. Details on such calculations and parametrization of the energy barrier are given in Section S9. We found that the diabatic, pure, nearest-neighbour CT pairs are about ~0.1 eV above the energy of localized FE states, which confirms that FE and CT excitations can potentially strongly mix and modify the optical properties of the aggregate. We also estimated a considerable exciton binding energy, $E_b$, in Y6 crystal of ~0.5 eV. This relatively large $E_b$ suggests against the facile generation of free charges starting from bound electron-hole CT pairs in pristine Y6 crystal. This point will be further investigated in the experimental section.

The spectra obtained from the diagonalization of the full Hamiltonian in Eq. 1 are shown in Fig. 3, without and with the inclusion of local vibronic coupling introduced using Eqs. 5-7 (see panel a and b, respectively). Interestingly, in both cases we note a shift of the main 0-0 peak of the aggregate towards lower energy compared to $E_{FE}$ (dashed black line). This additional red shift (of ~50 meV) acting on top of the STCS is a direct consequence of the interaction between FE and CT states. Specifically, the mixing between these states gives rise to hybrid FE-CT states which are pushed down in energy similarly to what was found by Spano et al. for perylene-based dyads, donor-acceptor-donor chromophores and other 1D systems.[30,39,40] In these cases, the interference between short-range photoinduced hole/electron transfer couplings and Coulomb interaction gives rise to "red-shifted HJ aggregates" depending on the phase of the interactions. In Fig. 3c,d, we report the character of the adiabatic states of Y6 crystal as a function of their energy (calculated with Eqs S11-S13) and we clearly show that FE and CT can mix to a different degree depending on the position in adiabatic energy space. Taking into account the associated red-shift now results in excellent agreement with the experiment absorption spectrum reported in Fig. 1c.

As shown in Fig. 3b, including the coupling to vibrations generates a vibronic shoulder in the spectrum, analogous to that in Fig. 2d where only FE states were considered. However, the relative intensities of the 0-0 and 0-1 vibronic replica dramatically change between the pure FE and the hybrid FE-CT model, the latter showing a much more prominent 0-1 band, in excellent agreement with the normalized experimental absorption on film (see Fig. 1c). By comparing the absolute intensity of the spectra without and with CT states, we can observe that the intensity of the 0-1 band remains roughly constant, but the intensity of the main 0-0 transition decreases in the former case (see Fig. S10). This is because the bright FE states at the onset of the absorption becomes strongly mixed with CT states (Fig. 3c,d) and lose part of their oscillator strength (compared to the pure FE case). Overall, this has the effect of reducing the ratio between the 0-0 and the 0-1 intensities when CT states are included in the Hamiltonian.

Finally, moving the attention to the blue edge of the spectrum (*i.e.*, ~690 nm) in Fig. 3c,d, we observe that the initially-dark CT states become slightly dipole allowed by modestly borrowing oscillator strength from vibronic FE states. This mixing is driven by the large density of pure CT states available at these energies and significantly broadens the aggregate spectrum with respect to the one in solution, thereby explaining also such an evident feature of the measured spectra shown in Fig. 1c. On the contrary, when only FE states are considered (Fig. 2d), the spectral broadening does not change going from solution to the aggregate.

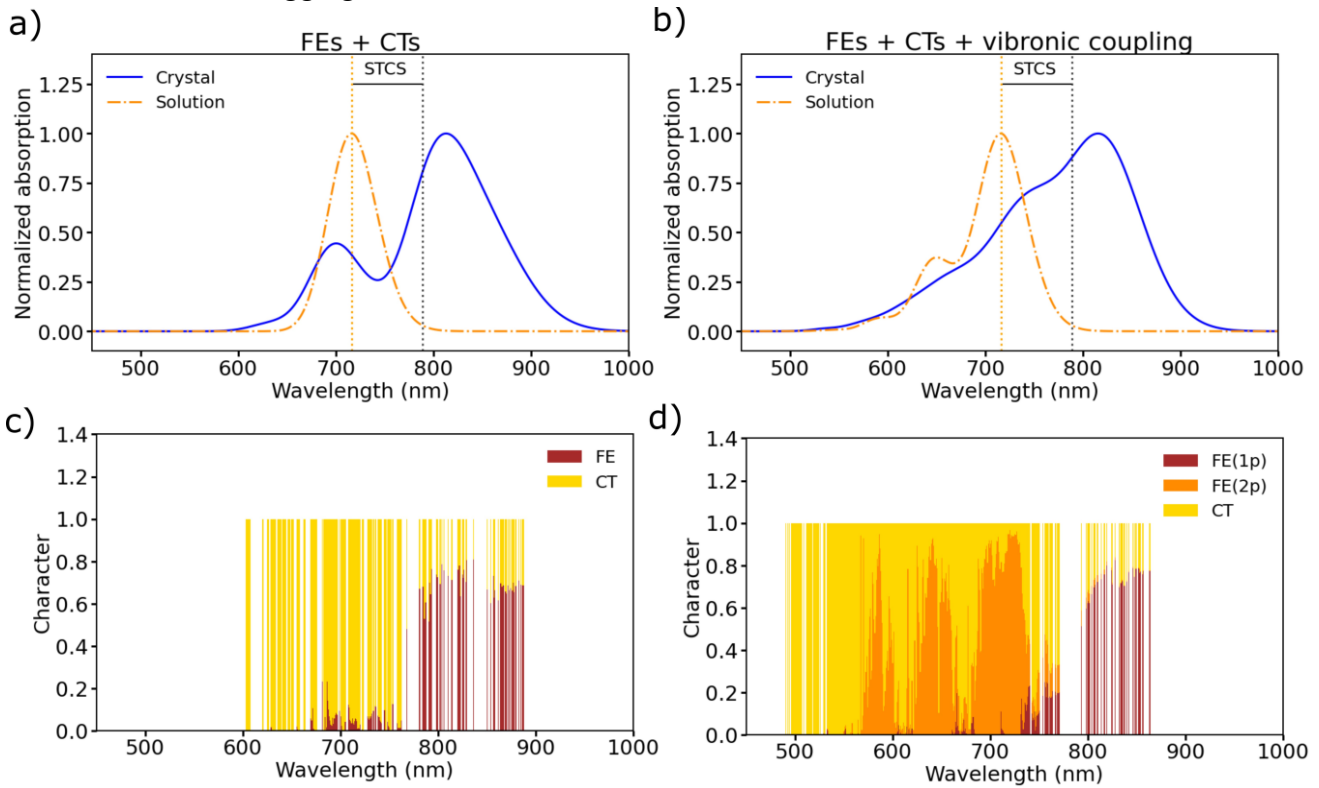

**Figure 3:** Panels **(a)** and **(b)** represent the computed and solution aggregate spectra without and with local vibronic coupling, respectively, obtained using the full Hamiltonian in Eq. 1. Panel b) shows the same data as Fig. 1d. The STCS is indicated. Dashed black vertical lines represent the FE energy ($E_{FE}^{(cry)}$), while the dashed orange ones represent the excitation energy in CF solution ($E_{S_0 \to S_1}^{(sol,CF)}$). Both values are obtained with BSE/MM as explained in the text. No shift is applied to the computed spectra. Panel **(c)** and **(d)** depict the character of the eigenstates of the Hamiltonian as a function of their energy and the related FE-CT mixing without and with the inclusion of vibronic coupling in the Hamiltonian, respectively. When the vibronic coupling is included FE one-particle and two-particles states (indicated with 1p and 2p) are depicted with different shades of orange (see Methods for details).

**Time resolved spectroscopy of Y6 thin-films**

To gain a deeper understanding of the excited-state dynamics in thin-film phase Y6 and how they fit in with our computational results, we turn to TAS. In Fig. 4a, we show the TAS spectrum of a neat Y6 film with excitation at 765 nm. Like the solution TAS measurement (reported in Fig. S13), the thin-film possesses a central ground state bleaching (GSB) in the region of 615–900 nm, flanked by two photo-induced absorption (PIA) bands; one centred at 550 nm and the other at 930 nm. However, it is informative to consider the form of the TAS spectrum in more depth. Specifically, the GSB exhibits a noticeable dip in the centre, approximately ranging from 700 to 825 nm, that does not match the reported steady-state absorption spectrum (Fig. 1c). As expanded on below, we therefore propose that this discrepancy in the spectral shape is caused by the presence of additional PIA underneath the GSB band.

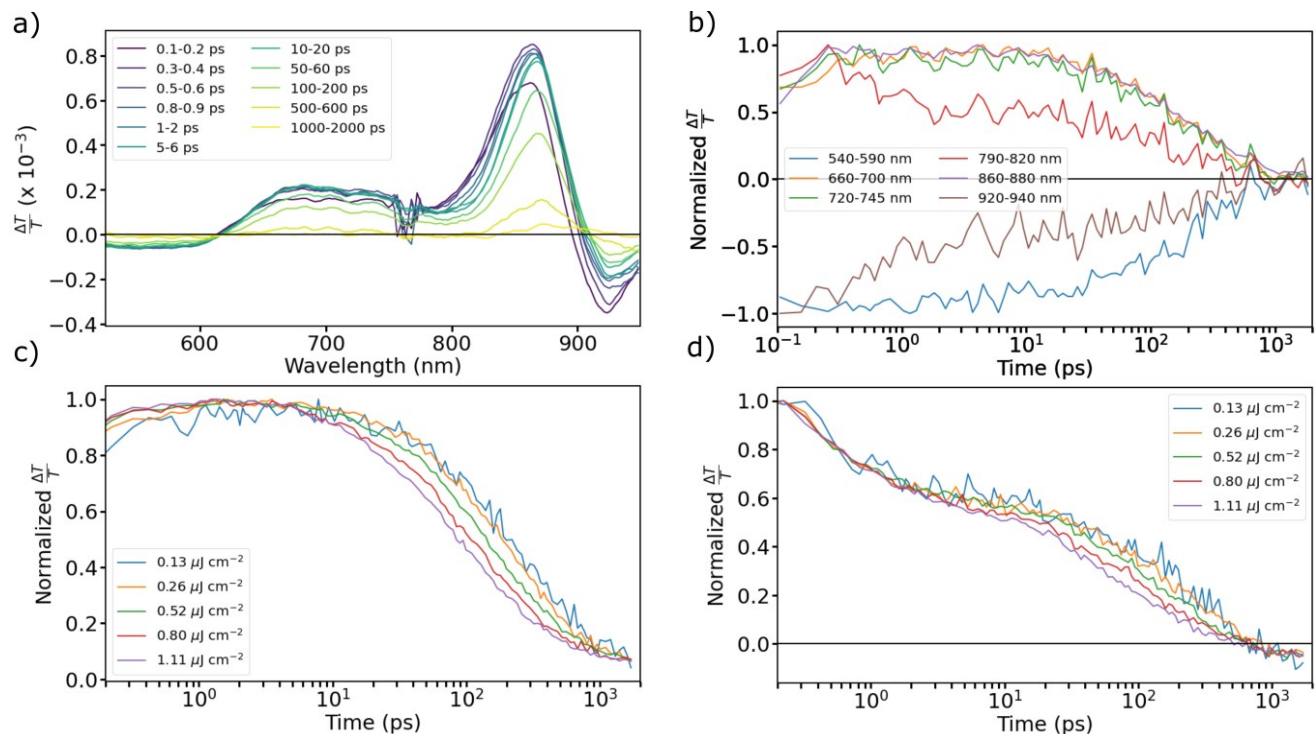

**Figure 4: (a)** Y6 neat (100wt%) film visible probe ps-TAS spectra. Note that the noise around ~765 nm is due to pump scatter in this region. **(b)** Kinetics normalized to maximum, fluence = 0.61 μJ cm$^{-2}$ ($\lambda_{pump}$ = 765 nm). Panels (c) and (d) depict Y6 neat film NIR probe ps-TAS fluence series for **(c)** 860-870 nm kinetic and **(d)** 920-960 nm kinetic ($\lambda_{pump}$ = 800 nm). In the latter two panels, kinetics are normalized to the maximum for each fluence.

Through comparison to TAS measurements on BHJ blends of Y6 with three commonly used electron donating polymers (Fig. S16), we assign this PIA band underlying the GSB to the (electron) polaron of Y6 (see Section S14 for discussion of the assignment). However, the PIA band centred at ~780 nm in the BHJ blends is more pronounced than in the neat Y6 film, to the extent that it dominates over the Y6 GSB feature located above it. Hence, in-line with previous reports,[7,9,11,12] there is an initial indication that polaronic-like states may also be present in the neat film of Y6, albeit formed with a lower yield than in the BHJ blends. To provide further evidence for the formation of polaronic states in Y6, Fig. 4b shows the normalized TAS kinetics taken from different spectral regions of the neat Y6 film. Most of the kinetics exhibit a single decay component. However, two kinetics in the 790–820 nm and 920–940 nm range are biphasic and display an additional fluence-independent sub-picosecond component. The PIA around 930 nm has previously been assigned to the singlet exciton state in Y6.[9] Though our calculations suggest that the initially photogenerated state will be specifically a FE-CT hybrid excitation, rather than a pure singlet exciton, we will refer to this state as the (singlet) exitonic state for convenience. Nonetheless, the decay of this state, which matches the growth of the Y6 polaronic PIA underneath the GSB, further confirms the conversion of excitonic states into polarons in neat Y6. Our ability to fully resolve this exciton to polaron conversion process is limited by the instrument response of ~200 fs for this TAS dataset. Thus, we have also performed additional TAS measurements with an ultrafast (<20 fs) excitation pulse (Fig. S19). Exploiting the superior time resolution, a kinetic taken from the singlet excitonic PIA between 920-940 nm reveals that its intensity reduces by ~60% within 2 ps after excitation, before plateauing (Fig. S20). This allows us to estimate that 60% of photogenerated excitations in a neat Y6 film form polaronic states, whilst 40% occupy the (singlet) excitonic states.

In-line with the observations of Wang et al., we find an additional PIA at 1550 nm in the neat Y6 film that is formed on the same timescales that the 930 nm singlet excitonic PIA is quenched on (see Fig. S17).[9] This implies that the origin of this 1550 nm PIA may also be polaronic. Curiously, when examining the dynamics of the BHJ blends, we find that the 1550 nm PIA is quenched by hole transfer from Y6 to the polymer (Fig S15). To further probe the origin of this absorption signature, we have performed spectroelectrochemical measurements on a neat Y6 film (Fig. S24). See details of the measurements in Methods. Under oxidation, we observe the formation of a new polaronic absorption band at 0.89 eV (1380 nm), close in energy to the 0.80 eV (1550 nm) PIA band in the TAS of neat Y6. We note that the different counterion ($PF_6^-$ in the spectroelectrochemistry vs $Y6^-$ in the TAS) could impact the hole polaron localisation and thus account for the small (<0.1 eV) energy difference of the measured transitions.[51] Therefore, as the 1550 nm PIA band shows a close match to the new infrared absorption feature in an oxidised Y6 film and is also quenched by hole transfer in the BHJ blends, we assign this feature to the hole polaron of Y6.

As an additional tool to probe the exciton to polaron conversion process, we have dispersed the Y6 molecules at 50 wt% in the non-interacting wide bandgap host material 3,3′-Di(9H-carbazol-9-yl)-1,1′-biphenyl (mCBP). TAS measurements reveal that diluting Y6 in the solid-state results in a slowing of the polaron formation (Fig. S15). Consequently, the spectral shape of the GSB at 0.1-0.2 ps much more closely resembles the steady-state absorption of neat Y6, before the polaronic PIA at 780 nm grows in over slightly longer (picosecond) timescales. By 10-20 ps, the spectral form in the mCBP dispersion has evolved to match that of neat Y6, indicating the process converting excitonic states into polarons is completed by this time. Therefore, it is apparent that the formation of the polaronic states in Y6 is extremely sensitive to the dispersion of the Y6 molecules and the resulting disruption in the formation of extended Y6 aggregates.

In Fig. S18, each kinetic taken from the TAS of Y6 is normalized to its value at 2.5 ps, and it can be observed that all the kinetics exhibit similar decay thereafter with a time constant of ~300 ps. Consequently, recovery to the ground state of most electronic excitations present after 2.5 ps is governed by the same process. The fluence dependence of the recombination kinetics is illustrated in Figs. 4c and 4d. Here, we only reach a regime where the excited state decay is fluence-independent at the very low fluences of 0.13 and 0.26 µJ cm$^{-2}$. This indicates that the recombination of electronic excitations in neat Y6 becomes bimolecular at relatively low excitation densities, consistent with the long excited state diffusion length reported for Y6.[52] Furthermore, we observe that the 860–870 nm kinetic taken from the peak of the GSB region reaches a plateau at 2 ns with approximately 8% of its normalized intensity remaining, independent of the fluence. At this point, we find no evidence for a significant remaining population of (singlet) excitonic or polaronic states on Y6, due to the absence of their PIAs at 930 nm, 780 nm, and 1550 nm (Figs. 4 and S21a). The only remaining PIA is that of the Y6 triplet exciton, found at ~1450 nm (Fig. S21c), which has previously been reported to decay via a mixture of mono-molecular triplet decay and triplet-triplet annihilation, depending on the triplet exciton density in the film.[53] As the triplet exciton yield is fluence independent, they cannot be formed from the states in neat Y6 that show a strong fluence dependence to their recombination rate, as this would be expected to reduce the triplet yield at higher excitation fluences. Thus, to account for this observation of a fluence-independent triplet exciton yield of ~8%, we propose the existence of a small but constant proportion of photogenerated singlet excitonic states that do not interconvert with the polaronic states, nor show significant bimolecular annihilation at the (low) excitation fluences measured here (see Fig. 5). We attribute this to the creation of singlet excitonic states on Y6 molecules located too far from a site where exciton to polaron transformation can occur (vide infra). These excitonic states may subsequently undergo intersystem crossing (ISC) to generate the long-lived triplet excitons and are responsible for the observed residual and fluence-independent GSB signal at 2 ns. We believe that this small fraction of relatively long-lived singlet excitonic states may also be responsible for the nanosecond photoluminescence decay observed in neat Y6 films,[22,23] which exceeds the ~300 ps primary excited state lifetime observed in the TAS.

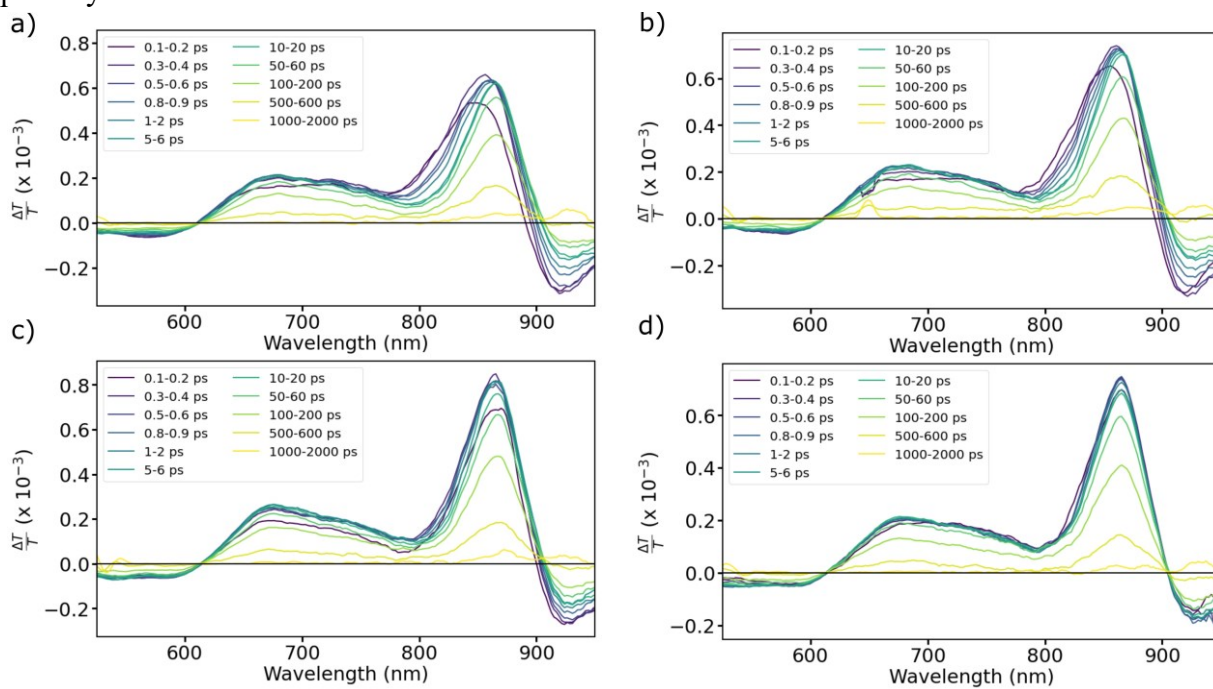

**Figure 5:** Excitation-dependent Y6 100 wt% thin-film visible probe ps-TAS spectra at (a) fluence = 2.26 µJ cm$^{-2}$ ($\lambda_{pump}$ = 500 nm), (b) fluence = 1.38 µJ cm$^{-2}$ ($\lambda_{pump}$ = 650 nm), (c) fluence = 0.98 µJ cm$^{-2}$ ($\lambda_{pump}$ = 870 nm), (d) fluence = 19.5 µJ cm$^{-2}$ ($\lambda_{pump}$ = 950 nm).

Our theoretical reproduction of the absorption spectrum for Y6 crystal suggests the presence of a continuum of hybridised FE and CT states, which is responsible for the significant broadening, and some of the red-shift, of the absorption spectrum when moving from solution to the thin-film environment. To test this hypothesis, we have investigated the pump wavelength dependence of the Y6 film TAS. We have selected four further pump wavelengths to excite different energetic regions in the Y6 absorption spectrum: 500 nm (2.48 eV), 650 nm (1.91 eV), 870 nm (1.43 eV), and 950 nm (1.31 eV), as shown in Fig. 5. Excitation at 500 and 650 nm is expected to pump higher lying FE-CT hybrid states, whilst 870 nm excitation will excite the lowest FE-CT transition. By contrast, sub band gap excitation at 950 nm is expected to pump the tail states present in Y6. We find that when pumping the higher lying FE-CT transitions at 500 and 650 nm, there is significant sub-picosecond spectral evolution. This is comparable to the TAS data shown in Fig. 5 with 765 nm (1.62 eV) excitation, where the evolution is particularly prominent in the regions associated with the Y6 polaron (~780 nm) and singlet exciton (~930 nm) PIAs, but even more pronounced. Curiously, the rapid spectral evolution is reduced, but still present, with 870 nm excitation. The spectral shape only becomes largely invariant at all measured time points when exciting in the absorption tail at 950 nm. This difference is further highlighted in Fig. S22, where the excitation energy-dependent spectra in the early (0.3-0.4 ps) and late (10-20 ps) time regimes are compared. In the early time regime, higher pump photon energies result in an enhanced intensity on the blue edge of the primary GSB peak around 800 nm and a more intense 930 nm PIA. In other words, the TAS spectrum shows the presence of more (singlet) excitonic states with a PIA at 930 nm and fewer polaronic states with a PIA at 780 nm. However, by 10-20 ps, the spectra become identical, indicating that the composition of the system is the same at this time regardless of the initial excitation wavelength. The amount of ultrafast evolution, which increases as the excitation photon energy rises, supports the calculation results of a dense continuum of hybridised FE and CT states through which the electronic excitation relaxes on sub-picosecond timescales, before eventually forming the final state with primarily more polaronic character (see Fig. 6). However, the continued presence of some spectral evolution with 870 nm excitation, which disappears with 950 nm excitation, suggests that the state with polaronic PIA signatures is not the state formed directly when exciting the lowest FE-CT transition of Y6. This points to the presence of additional, lower energy, polaronic states in Y6 films, distinct to the FE-CT manifold, that can only be directly accessed by sub band gap excitation at 950 nm.

Finally, it is necessary to address the unusual behaviour of the (singlet) excitonic PIA at 930 nm in the neat Y6 film. Specifically, two observations: (i) why the singlet excitonic states are not fully converted into the lower energy polaronic states, as seen by the PIA band at 930 nm retaining ~40% of its initial intensity beyond the timescales (>2 ps) of the polaronic state formation (Fig. S22a); and (ii) why the PIA signature of the excitonic states is present when directly exciting the polaronic states below the band gap at 950 nm. To explain these observations, we compare the kinetics taken from the singlet exciton PIA at 925-935 nm for the pump wavelength series in Fig. 5. For 500, 650, 765, and 870 nm excitation, there is a clear partial sub-picosecond quenching of the excitonic PIA as it is converted into the polaronic state. However, with 950 nm excitation, we note that the excitonic PIA instead shows a slight growth within the first picosecond after excitation, the rise of which we are likely unable to fully resolve due to the ~200 fs instrument response of this TAS measurement. As lowering the photon energy by 130 meV is enough to move from exciting the lowest FE-CT hybrid states to the sub band gap polaronic states, they are both clearly located close in energy. Thus, we conclude that the excitonic and polaronic states in Y6 are in a dynamic equilibrium, which is why their PIA signatures co-exist and there is evidence for their interconversion in the TAS. A dynamic equilibrium between excitonic and polaronic states can also explain the largely identical decay kinetics seen across all spectral regions after 2.5 ps in Fig. S23, as both manifolds can share a common dominant (non-radiative) recombination pathway, likely via the polaronic state manifold. For convenience, we have summarised the proposed states and their photophysics following optical excitation in Fig. 6. This revolves around the presence of two distinct FE-CT hybrid state manifolds, one of which can interact with the polaron pair (PP; vide infra) states, whilst the other cannot. The former manifold of 'interacting' FE-CT states is responsible for the polaron pair formation and the majority of the (non-radiative) recombination, whilst the latter 'isolated' FE-CT states mediates the nanosecond fluorescence and fluence-invariant triplet exciton yield via ISC also observed in neat Y6 films.

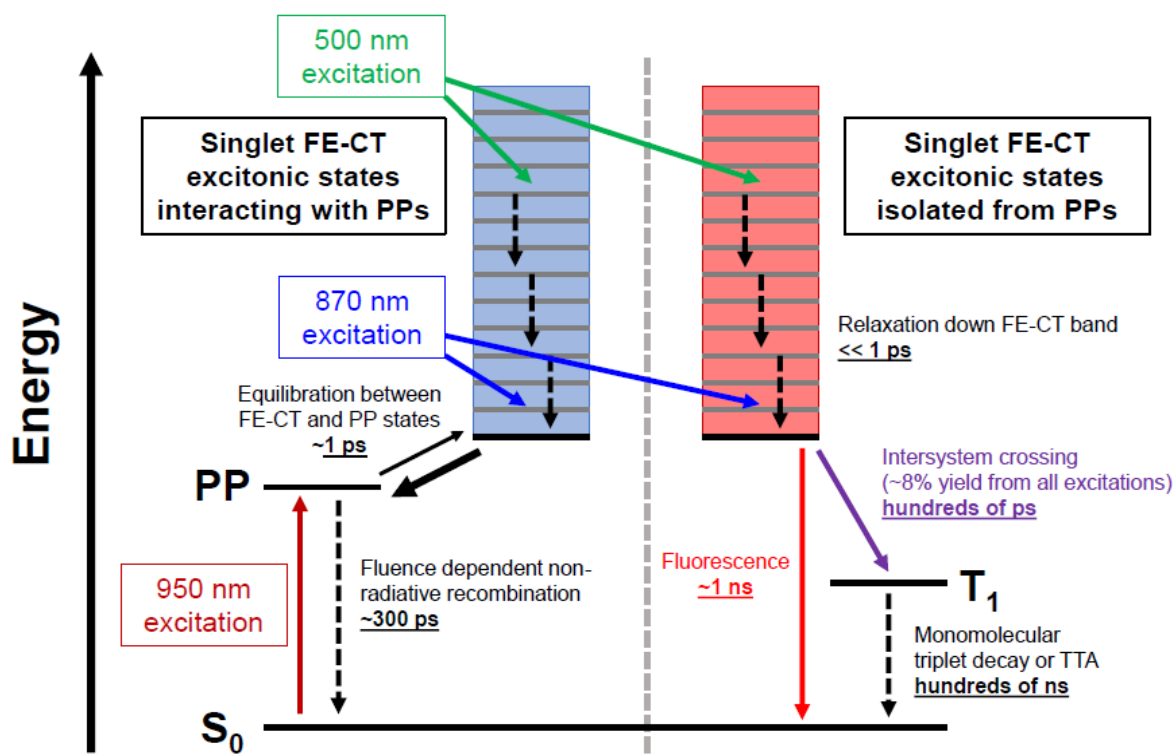

**Figure 6:** A schematic representation of the excited state dynamics of Y6 thin-film. To explain the complex Y6 photophysics, we propose presence of two distinct Frenkel exciton-charge transfer (FE-CT) hybrid excitation manifolds in the neat Y6 film: one that can interact with the polaron pair (PP) states observed experimentally, and one that cannot. The timescales of the relevant electronic processes occurring in Y6 is also given on the diagram.

**The nature of the polaronic states in Y6 films**

Our experimental results thus far indicate that most photo-generated electronic excitations in Y6 eventually form states with polaronic signatures, independent of the excitation photon energy. However, we note that such a low-lying polaronic state is not captured by our computational models based on the crystal structure data of Y6. Thus, when combined with our experimental evidence pointing to the presence of a sub band gap polaronic state, we conclude that polaron formation is not an intrinsic property of the FE-CT hybrid excitations of Y6 aggregates, but rather results from regions of the 'real world' Y6 film that are not present in the 'perfect' single crystal environment. This could include defect sites in the molecular packing motif or boundaries between crystalline domains in the Y6 film. For example, in a recent paper on single-component α-sexithiophene device that gave a power conversion efficiency of 2.9%, a plausible explanation for free polaron generation was proposed based on electrostatics at the interface between crystalline domains possessing distinct orientations.[20] It is possible that a similar electrostatic effects could also be present in neat Y6 films, as indicated in a recent publication by Fu et al.[17]

As there is currently significant debate on this topic,[7,11,22] it is *important* to clarify whether these polarons have a significant separation and can be considered free charge carriers (defined as when the electron and hole separation exceeds their Coulomb capture radius of ~5 nm[21]) or whether they represent more tightly bound polaron pair states, potentially located on neighbouring Y6 molecules. On this point, we first note that the lifetime of the polaronic states in the TAS of neat Y6 films is on the order of ~300 ps in the fluence-independent regime. Given the short, sub-nanosecond lifetime, we consider that this already provides good evidence that the polarons do not achieve long range charge separation; the free charge carrier lifetimes reported in organic solar cell blends of PM6:Y6 are on the order of microseconds to tens of microseconds in TAS measurements with comparable, or slightly higher, excitation fluences.[12] Thus, it seems implausible that neat Y6 should also generate polarons that achieve a separation exceeding the Coulomb capture radius of ~5 nm,[21] yet decay four to five orders of magnitude more quickly than in the PM6:Y6 BHJ. Even when considering that the BHJ morphology can assist with keeping the electrons and holes separate.

To better understand the nature of the polaronic states in neat Y6 films, we turn to magnetic resonance techniques (see Methods). Magnetic resonance spectroscopies are uniquely sensitive to the presence of unpaired spins in a sample and are thus well placed to probe electronic excitations with a non-zero spin, such as polarons, (triplet) charge transfer states, and molecular triplet excitons. We begin by exploring Y6 with transient electron paramagnetic resonance (trEPR) spectroscopy. Fig. 7a shows the trEPR spectrum of neat Y6 film. The full-field (FF) spectrum (280 - 420 mT) corresponds to $\Delta m_S = \pm 1$ transitions between triplet sublevels, whilst the width of the FF signal is a measure of the Zero-Field Splitting (ZFS) parameter $D$, which in organic semiconductors is determined by the dipolar interaction and thus correlated to the inter-spin distance $r$ ($D \sim r^{-3}$).[54] The $D$ value is given by $D$ = 940 MHz characteristic for Y6 molecular triplet excitons,[55,56] whilst the *eeeaaa* polarization pattern (*e* = emission, *a* = absorption) is typical for triplet excitons formed via ISC with a zero-field population of $p_z$, $p_y$, $p_x$ = 0, 0.66, 0.34. However, it is pertinent to compare Y6 to other organic semiconductor materials that are known to form a small quantity of free polarons in neat films, such as conjugated polymers.[57] One example is the commonly used OPV electron donor polymer PBDB-T,[58] for which the trEPR spectrum is shown in the inset of Fig. 7a. As well as the broad molecular triplet exciton feature, we also observe the presence of a narrow feature at ~336 mT. The small width of this feature and the presence of a single peak, as opposed to a more complex polarisation pattern with several emissive and absorptive peaks typical of a charge transfer state,[59] indicates that it originates from free polarons. By contrast, this signature is absent in Y6. Whilst it is worth noting that the time resolution of trEPR is on the order of hundreds of nanoseconds and the TAS measurements suggests the majority of excited state recombination in Y6 occurs sub-nanosecond, the absence of a polaron signal further suggests that the formation of long-lived free polarons in Y6 is a negligible pathway following photoexcitation. An even more sensitive magnetic resonance method with optical detection and cw-light excitation is photoluminescence-detected magnetic resonance (PLDMR). Fig. 7b shows the PLDMR spectrum of a spin-coated Y6 film. The same Y6 molecular triplet exciton observed in the trEPR is also seen here in the pronounced spectral 'wings' at ~303 and 386 mT. The details of the molecular triplet exciton in the PLDMR spectrum of Y6, including its link to the strong molecular ordering face on to the substrate of Y6, has been discussed previously.[56] However, we also note the presence of a narrower feature around $g = 2$. The linewidth of this feature in neat Y6 is 2.5 mT (see EasySpin simulations, Fig. S25 and Table S9), which is comparable to reports of CT states located at the donor:acceptor interface in OPV blends, such as PM6:Y6 or PBDB-T:Y6.[56] By contrast, the PLDMR of neat polymers that showed free polaron features in trEPR, such as PBDB-T, have a smaller linewidth of 1.0 – 1.4 mT (as $D \sim r^{-3}$, a small change in linewidth correlates to a significant difference in the interspin distance, $r$).[56] Note that for weakly-interacting spins, the two transitions probed in PLDMR between the $m_s = 0$ and $m_s \pm 1$ sublevels of the (triplet) CT state (equivalent to a polaron pair state), will merge and appear as a single inhomogeneously broadened Gaussian curve.[56] Thus, when considered together with the absorption signatures of polaronic states in the TAS and the absence of free polarons in the trEPR, the PLDMR results suggest that the polaronic states present in neat Y6 may be polaron pairs with a small intermolecular separation, likely located on nearest-neighbour Y6 molecules. However, further PLDMR studies are required to identify the precise nature of the polaronic states in Y6 and their paramagnetic properties, which are beyond the scope of this work. Based on the present evidence, whilst it is plausible that these polaron pair states may possess a reduced binding energy that renders them more susceptible to separation in a single-component Y6 OPV device, potentially at the interface with the hole and/or electron transport layers or assisted by an electric field applied across the device, we conclude that free polarons are likely not the primary photoexcitation in neat Y6 films.

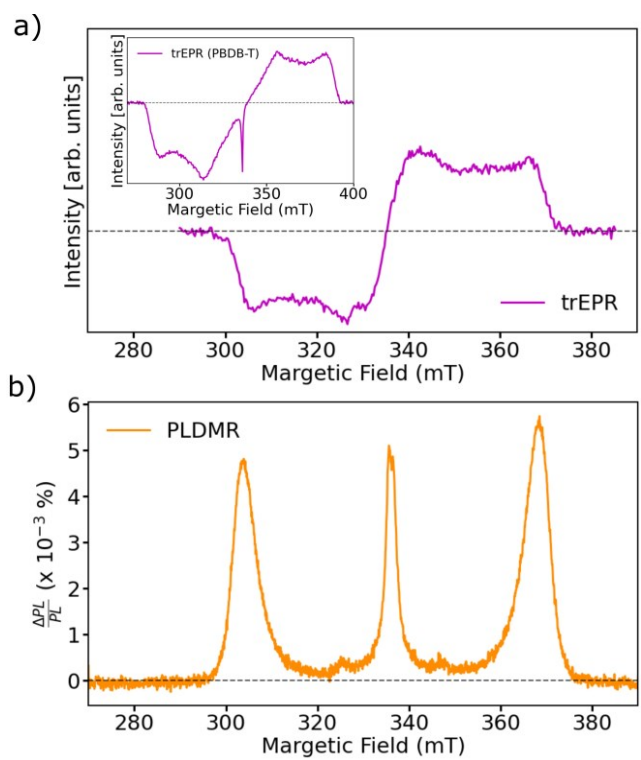

**Figure 7:** Magnetic resonance measurements on neat Y6 films. **(a)** trEPR spectrum for neat Y6 film (inset: trEPR spectrum for neat PBDB-T film). **(b)** PLDMR spectrum of neat Y6 film **(a)** trEPR spectrum for a neat Y6 film (inset: trEPR spectrum for a neat PBDB-T film). All measurements were obtained using $\lambda_{pump}$ = 532 nm and $\lambda_{pump}$ = 473 nm and were performed at T = 10 K.

## Conclusions

We have shown that high-quality electronic structure calculations combined with a reliable description of the environmental response, both in solution and in the solid state, account for the experimentally observed spectral changes upon crystallization in the archetypic NFA Y6. Notably, this degree of accuracy is only possible when the energetics of the CT excitations and their mixing with other states is considered. We demonstrated in this way that two major effects are responsible for the observed red-shift going from Y6 solution to its thin-film-phase: (i) strong mixing between closely spaced diabatic FE and intermolecular CT states (with an energy off-set for the closest contact pair being ~0.1 eV in Y6); (ii) a sizable solution-to-crystal shift going from chloroform solution to solid-state Y6, which we calculate to be about 0.16 eV. We have shown that the steady-state optical features of Y6 upon aggregation are strongly influenced by the strength of the intermolecular interactions, for instance the shape and relative intensity of the vibronic shoulder compared to the main absorption band as well as the broadening of the Y6 thin-film spectrum compared to the solution counterpart are directly a consequence of the formation of hybrid FE-CT states.

We have provided experimental evidence for the proposed hybridisation of FE and CT states in Y6 by probing the pump wavelength dependence of the excited state dynamics using TAS. At higher photon energies, there is an increase in the sub-picosecond relaxation effects in the Y6 film, which we attribute to the relaxation of the electronic excitations down the dense manifold of FE-CT hybrid states predicted computationally. However, TAS also reveals the existence of a polaronic state below the lowest FE-CT state, which can be directly accessed by excitation below the band gap of Y6 at 950 nm. We further find evidence that the excitonic and polaronic states are in a dynamic equilibrium, as exciting the polaronic states at 950 nm leads to the delayed formation of (singlet) excitonic signatures in the TAS. However, the short (~300 ps) lifetime of these polaronic states, combined with the absence of any features expected for free polarons in magnetic resonance spectroscopy measurements, suggests that most of these states are bound polaron-pairs, likely located on neighbouring Y6 molecules. The absence of such a polaron pair state in our high-quality electronic structure calculations further implies that the presence of polaronic states is related to factors not captured by the 'perfect' single crystal environment, such as defect sites or interfaces between crystalline domains.

Considerations made in this work, which led us to unravel important electronic and spectroscopic features of Y6, likely apply more broadly to other NFAs within the Y-family. For instance, due to the specific and complex crystal packing, the effects of textbook H- and J-like excitonic interactions on the spectral features of the solid-state phase of these NFAs are expected to cancel out. This most probably leads to other factors (FE and CT states hybridization, environmental effects) being responsible for the peculiar electronic properties and photophysics of Y-family systems, as we have found here for the archetypic Y6. By unravelling the mystery of what makes Y6 special, our work provides a blueprint for the intelligent optimization of this material class to enhance its properties for OPV applications. Specifically, we propose that the field should focus on understanding the relationship between solid-state packing motifs, interaction networks, delocalization of exciton and charges and environmental effects. For example, by examining the impact of different chemical modifications to the molecular backbone of Y-series materials on the solid-state morphology, electronic properties, and excited state dynamics. Moving forward, explicit time-dependent propagation of coupled electron-nuclear motion through non-adiabatic molecular dynamic simulations[18,60,61] on realistic and more complex nano-scale microstructures could help us achieve this goal. By leveraging these structure-properties relationships, the optical and electronic properties of the NFA can be effectively controlled, potentially yielding new Y-series NFA materials that can drive power conversion efficiencies toward the elusive 20% milestone.

## Methods
**Frenkel-Holstein Hamiltonian.**

To model optical properties and electronic states of tightly packed molecular aggregates with excited states of different characters, *i.e.*, FE and CT states, in the presence of local vibronic coupling to effective vibrational modes for hole, electron and excitonic states, we utilized the following Frenkel-Holstein-type Hamiltonian written as:

$$H = H_{\text{FE}} + H_{\text{CT}} + H_{\text{FE-CT}} + H_{\text{N}} + H_{\text{FE-N}} + H_{\text{CT-N}} \qquad (1)$$

$H_{\text{FE}}$ represents the Frenkel-exciton block which describes the interactions between bound electron-hole pair residing on the same molecular site $k$ and can be written as:

$$H_{\text{FE}} = \sum_{k}\left(E_{S_0 \to S_1}^{(\text{sol})} + \Delta_{0-0}\right)|e_k\rangle\langle e_k| + \sum_{k,l} V_{kl}|e_k\rangle\langle e_l|. \qquad (2)$$

Here, the energy of a local Frenkel exciton ($|e_k\rangle$), in the crystal is given by $E_{\text{FE}}^{(\text{cry})} = E_{S_0 \to S_1}^{(\text{sol})} + \Delta_{0-0}$, where $E_{S_0 \to S_1}^{(\text{sol})}$ is the energy of the electronic transition in solution and $\Delta_{0-0}$ is the solution-to-crystal shift (which is a negative shift). The off-diagonal elements of this block are the excitonic couplings, $V_{kl}$, which describe the interactions between tightly bound electron-hole pair sitting either on molecules $k$ or $l$ and allow for excitation energy transfer between the two. These couplings are dominated by the Coulombic interaction between the transition densities of the two molecules. The magnitude of such an interaction scale as $r^{-3}$, where $r$ is the intermolecular distance. We refer to Section S6 for a detailed explanation of how all these terms are calculated in practice using a combination of TDDFT and accurate BSE calculations.

The second term $H_{\text{CT}}$ represents the charge transfer Hamiltonian which describe the interactions between charge transfer excitons. This is written as:

$$\begin{aligned}H_{\text{CT}} = &\sum_{k,s\neq 0}\left(E_{\text{CT}}(r_s)\right)|a_k, c_{k+s}\rangle\langle a_k, c_{k+s}| \\ &+ \sum_{k,s\neq 0, s'\neq 0} t_{\text{h}}(s-s')\,|a_k, c_{k+s}\rangle\langle a_k, c_{k+s'}| \\ &+ \sum_{k,s\neq 0, s'\neq 0} t_{\text{e}}(s-s')\,|a_{k+s}, c_k\rangle\langle a_{k+s'}, c_k|\end{aligned} \qquad (3)$$

where the $|a_k, c_{k+s}\rangle$ denotes the charge transfer exciton where the cationic (anionic state) is located on the site $k$, while the hole (cationic state) is on molecule $l$ located at distance $s$ from the first molecule. The vectors $s$ and $s'$ denote the electron/hole separation (in lattice units) and are restricted to nonzero values. When $s = 0$ the electron and hole are located on the same molecule resulting in a FE excitation. $E_{\text{CT}}(r_s)$ is the energy of the charge transfer state and it is determined in this work as described in Section S9. This term is defined by the Coulomb binding energy of hole and electron at a given distance $r_s$. $E_{\text{CT}}(r_s)$ is therefore the smallest when the interacting hole and electron reside on the closest crystal pair and the energy difference of this state with respect to $E_{S_0 \to S_1}^{(\text{cry})}$ determines the energy offset ($\Delta_{\text{offset}}$) between FE and CT states.

CT excitons couple to one another via electron and hole transfer integrals, $t_e$ and $t_h$, respectively. The charge transfer integrals are a result of the interactions between either the HOMO orbitals or the LUMO orbitals of two interacting chromophores, and they are influenced by the extent of spatial overlap between them. Consequently, the coupling decreases exponentially as the distance between the chromophores increases, and it is typically noteworthy only for closely interacting molecules (*e.g.*, nearest neighbours), see Section S7 for details of the actual computation of these terms. The mixing between FE and CT excitons is given by the Hamiltonian block $H_{\text{FE-CT}}$ which is written as:

$$H_{\text{FE-CT}} = \sum_{k,s\neq 0} D_h(s)|e_k\rangle\langle a_k, c_{k+s}| + h.c.$$
$$+ \sum_{k,s\neq 0} D_e(s)|e_k\rangle\langle a_{k+s}, c_k| + h.c. \quad (4)$$

where $D_e(s)$ and $D_h(s)$ are the photoinduced electron and hole transfer interactions, respectively, which couple FE and CT states, and where h.c. stands for the Hermitian conjugate. This Hamiltonian block is crucial because it allows dark CT states to borrow oscillator strength from bright FE. This, in turn, significantly impacts the optical properties of the aggregate. It is worth noticing that as carefully described in Ref.[62], $D_e(s)$ and $D_h(s)$ are not strictly the same as $t_e$ and $t_h$, since additional exchange-like terms appear in the expression of the former couplings. However, the leading contribution in the calculation of $D_e(s)$ and $D_h(s)$ is related to the one-electron terms (Fock matrix elements) involving the interaction of LUMO orbitals for the former and HOMO orbitals for the latter as it is the case for $t_e$ and $t_h$. Consequently, these terms decay exponentially with the distance between donor and acceptor sites as well.

Nuclear degrees of freedoms are also accounted for in our Hamiltonian. The electronic excitations (*i.e.*, the formation of anionic, cationic and excitonic states) are characterized by a strong coupling with high-frequency vibrational modes (which are represented here by a single effective mode with effective energy $\hbar\omega_{\text{eff}}$, to make the problem computationally feasible). This vibronic interaction is responsible for the pronounced vibronic progression observed in the spectra. $H_N$ describes the nuclear energy and it is written as:

$$H_N = \hbar\omega_{\text{eff}} \sum_k b_k^\dagger b_k \quad (5)$$

where $b_k^\dagger$ and $b_k$ are the common creation and annihilation operators associated with a quantum harmonic oscillator. Here, we have assumed that both excited and ground states to have the same curvature and we have also omitted the zero-point energy, which is uniform for all molecules. This exclusion does not impact our findings since we are focusing on the energy differences between the eigenstates.

Finally, the coupling between electron and nuclei is given by $H_{\text{FE-N}}$ and $H_{\text{CT-N}}$ terms written as:

$$H_{\text{FE-N}} = \hbar\omega_{\text{eff}}\sqrt{S_e} \sum_k (b_k^\dagger + b_k + \sqrt{S_e}) |e_k\rangle\langle e_k| \quad (6)$$

and

$$\begin{aligned}H_{\text{CT-N}} &= \hbar\omega_{\text{eff}}\sqrt{S_a} \sum_{k,s\neq 0} (b_k^\dagger + b_k + \sqrt{S_a}) |a_k, c_{k+s}\rangle\langle a_k, c_{k+s}| \\ &+ \hbar\omega_{\text{eff}}\sqrt{S_c} \sum_{k,s\neq 0} (b_{k+s}^\dagger + b_{k+s} + \sqrt{S_c}) |a, c_{k+s}\rangle\langle a_k, c_{k+s}|\end{aligned} \quad (7)$$

Where $S_{e,c,a}$ are the Huang-Rhys factors of the excitonic, cationic and anionic states, respectively, related to the displacement ($d_{e,c,a}$) of the excited state potential (*i.e.*, anionic, cationic or excitonic) with respect to the ground state. As described in Section S3, $S_{e,c,a} = \frac{\lambda_{e,c,a}^{\text{rel}}}{\hbar\omega_{\text{eff}}} = \frac{m\omega_{\text{eff}} d_{e,c,a}^2}{2\hbar}$. In Eq. 7 the first summation accounts for vibronic coupling to cationic states while the second term accounts for the vibronic coupling to anionic states.

The Hamiltonian matrix is expressed in a multi-particle basis set, which is commonly truncated to two particle states referred to as the two particle approximation (TPA), and numerically diagonalized.[31,32] Such a basis set has been extensively used by Spano *et al.*[31,32] to represent the low energy eigenstates of the Hamiltonian in the regimes of weak and intermediate electronic coupling. In this basis, the eigenstate *j* of the Hamiltonian can be written as:

$$\begin{aligned}|\Psi^{(j)}\rangle &= \sum_k \sum_{\tilde{v}=0}^{v_{\max}} c_{k,\tilde{v}}^{(j)} |e_k, \tilde{v}_k\rangle + \sum_k \sum_{l\neq k} \sum_{\tilde{v}=0}^{v_{\max}-1} \sum_{v=1}^{v_{\max}-\tilde{v}} c_{k,\tilde{v},l,v}^{(j)} |e_k, \tilde{v}_k; g_l, v_l\rangle \\ &+ \sum_k \sum_{l\neq k} \sum_{\tilde{v}^+=0}^{v_{\max}} \sum_{\tilde{v}^-=0}^{v_{\max}} d_{k,\tilde{v}^-,l,\tilde{v}^+}^{(j)} |a_k, \tilde{v}_k^-; c_l, \tilde{v}_l^+\rangle\end{aligned} \quad (8)$$

Where the first two terms correspond to one- and two-particle FE excitations, respectively, with $\tilde{v}$ and $v$ representing the vibrational energy levels in the shifted excited state potential and ground state potential respectively. While the last term introduces two-particle charge transfer states in which $\tilde{v}^+$, vibrational quanta reside on the cation and $\tilde{v}^-$ vibrational quanta reside on the anion. A detail description of the computation of the different parameters is provided in SI along with the related level of theory used for each parameter.

**General simulation details.**
Electronic structure calculations have been performed for the triclinic Y6 unit cell reported by Xiao *et al.*[36] Due to the very large thermal ellipsoids of the published structure, bond lengths and angles of individual molecules have been optimized (as described in Section S1). Y6 alkyl side chains have been replaced by methyl groups in all TDDFT and many-body calculations.

The polarizable continuum model (PCM) has been used to describe Y6 in solution,[24] using the geometries optimized in each specific solvent. Solid-state calculations considered electrostatic and dielectric embedding in the infinite bulk limit, described at the atomistic level.[24,25] The classical (MM) environment has been described with the induced-dipole model implemented in the MESCal code,[29] parameterized with polarizability tensor and atomic charges (ESP scheme) computed with (TD)DFT. The classical dielectric response of alkyl side chains, not provided in the experimental crystal structure [36], has been modelled with a grid of polarizable points occupying the corresponding cavity and bearing the polarizability of the replaced alkyl chains. The static dielectric tensor of the Y6 triclinic unit cell, built as in Section S1, has been computed with a microelectrostatic (ME) model by applying a uniform electric field along the three Cartesian directions, as explained in Ref [29] It has been previously shown that ME calculations, performed with the MESCal code, provide accurate values of the susceptibility tensor for crystal and molecules films. We refer the interested reader to Ref.[63]

The Frenkel-Holstein Hamiltonian in Eq. 1 has been constructed for Y6 crystal using a combination of embedded many-body GW/BSE/MM (for the site-energies) and TDDFT (for the electronic interactions) as described in the main text and further detailed below. The converged Hamiltonian is constructed for 3x3x3 cell sizes and using 3 vibrational quanta for all electronic excitations (FE and CT states). Periodic boundary conditions are taken into account using a minimum image convention for all the electronic interactions. Convergence with respect to different parameters used in the Hamiltonian is presented in Fig. S9.

**Embedded many-body GW/BSE and Time dependent density functional theory.**
Embedded many-body GW and BSE calculations have been performed with the FIESTA and beDeft codes,[64,65] adopting the formalism described in the original papers.[24,25] Kohn-Sham orbitals have been computed at the PBE0/6-311G(d) level of theory with the ORCA package.[66] Many-body calculations employed resolution-of-the-identity (RI-V) techniques,[67] using the universal Weigend Coulomb fitting basis set.[68] Quasiparticle energies have been computed with a self-consistent scheme on eigenvalues (evGW), correcting one (two) occupied and unoccupied levels for a Y6 molecule (dimer). Neutral singlet excitations have been obtained at the full BSE level, using an active space constructed by occupied-virtual transitions between states within 15 eV from the corresponding band edge. Solid-state BSE/GW/MM calculations have been performed for the different QM/MM partitioning schemes, namely considering as QM region: (i) each of the two symmetry-unique Y6 molecules in the crystal (results in Table S4 and S5); (ii) eight Y6 dimers chosen among the closest distance pairs (results in Table S7). The latter have been used to parameterize the model Hamiltonian for CT excited states energy of the crystal as explained in Section S9.

The electronic couplings, vibrational analysis and other parameters are computed using TDDFT. Specifically, we used an optimally-tuned (OT) range-separated hybrid (RSH) functional, unless stated otherwise. The tuning procedure adopted in this work with the LC-ωhPBE functional and the 6-311G(d,p) basis set is described in Section S2 and the OT ω value was found at 0.101 Bohr$^{-1}$ for the Y6 molecule.[69] With such a level of theory, excitonic couplings are computed using the multi-state diabatization fragment excitation energy difference-fragment charge difference (MS-FED-FCD) approach.[49,70,71] The electronic charge transfer couplings were calculated instead with the projection operator diabatization approach (POD)[72–74]. Both these approaches are briefly described in Section S6 and S7.

**Optical Steady-State Absorption**

The steady-state absorption spectra of Y6 in chloroform and toluene solution has been recorded using a high-performance quartz cuvette (Hellma Analytics). The absorption spectrum has been determined via a transmission measurement using an UV-Vis-NIR spectrometer (Agilent Technologies Cary 5000 Series). The concentration of Y6 has been determined by diluting the solution so that the signal was not saturated as well as invariant under consecutive dilutions.

The film spectra have been obtained using a combination of reflection-transmission measurements using an integrating sphere, and reverse transfer matrix modelling approach developed by Kerremans *et al.* to obtain the optical constants of the films.[37] Three separate films with different thicknesses have been fabricated from chloroform solvent. This was done by spin casting the material on borosilicate glass substrates at 2000 rpm. All films were annealed at 120 °C. The different thicknesses have been obtained by dissolving the Y6 in solutions with different concentrations. The processing parameters and sample thicknesses are listed in Table S8, and the reverse transfer matrix modelling described more in details in Section S12. Sample thicknesses have been measured using a profilometer (Bruker DekTakXT).

**Transient Absorption Spectroscopy**

TAS has been performed on either one of two experimental setups. The femtosecond TAS in the IR region (1200 – 1600 nm) has been performed on a setup powered using a commercially available Ti:sapphire amplifier (Spectra Physics Solstice Ace). The amplifier operates at 1 kHz and generates 100 fs pulses centred at 800 nm with an output of 7 W. A portion of the laser fundamental was used for sample excitation at 800 nm. The broadband infrared probe pulse used in the femtosecond TAS measurements is provided by a home built non-colinear optical parametric amplifier (NOPA). The probe pulses are collected with an InGaAs dual-line array detector (Hamamatsu G11608-512DA), driven and read out by a custom-built board from Stresing Entwicklungsbüro. The probe beam has been split into two identical beams by a 50/50 beamsplitter. This allowed for the use of a second reference beam which also passes through the sample but does not interact with the pump. The role of the reference was to correct for any shot-to-shot fluctuations in the probe that would otherwise greatly increase the structured noise in our experiments.

For the 500 – 950 nm continuous probe region TAS, a Yb amplifier (PHAROS, Light Conversion), operating at 38 kHz and generating 200 fs pulses centred at 1030 nm with an output of 14.5 W has been used. The ~200 fs pump pulse was provided by an optical parametric amplifier (Light Conversion ORPHEUS). The probe is provided by a white light supercontinuum generated in a YAG crystal from a small amount of the 1030 nm fundamental. After passing through the sample, the probe is imaged using a Si photodiode array (Hamamatsu S11490).

**Magnetic Resonance Spectroscopies**

Photoluminescence-detected magnetic resonance (PLDMR) and transient electron paramagnetic resonance (trEPR) experiments were carried out with a modified X-band spectrometer (Bruker E300) equipped with a continuous-flow helium cryostat (Oxford ESR 900) and a microwave cavity (Bruker ER4104OR, 9.43 GHz) with optical access. All measurements have been performed at T = 10 K.

For PLDMR, microwaves have been generated with a microwave signal generator (Anritsu MG3694C), amplified to 3 W (microsemi) and guided into the cavity. Optical irradiation has been performed with a 473 nm continuous wave laser (Cobolt). PL has been detected with a silicon photodiode (Hamamatsu S2281) on the opposite opening of the cavity, using a 561 nm longpass filter to reject the excitation light. The PL signal has been amplified by a current/voltage amplifier (Femto DHPCA-100) and recorded by lock-in detector (Ametek SR 7230), referenced by on-off-modulating the microwaves with 517 Hz.

For trEPR, pulsed optical excitation has been performed with a Nd:YAG laser (Continuum Minilite II) with 532 nm; pulse length of 5 ns; 15 Hz repetition rate; 2 mJ per pulse. Microwaves have been generated and detected with a microwave bridge (Bruker ER047MRP). Measurements have been performed with 20 dB attenuation (2 mW). A voltage amplifier (FEMTO DHPVA-200) and a digitizer card (GaGe Razor Express 1642 CompuScope) have been used for transient recording. The time resolution is limited to ~100 ns by the cavity Q factor of around 2800. By sweeping the magnetic field, two-dimensional data sets are recorded, where trEPR spectra are averaged from 0.5 – 1.5 μs after laser excitation.

**Spectroelectrochemistry**
For the spectroelectrochemical samples, a 16 mg/mL solution of Y6 dissolved in chloroform was prepared and then stirred overnight at 45 °C under an inert nitrogen atmosphere. ITO coated glass substrates were sonicated in deionized water, acetone, and then isopropyl alcohol for 10 min each, and then were plasma ashed with $O_2$ plasma for 10 min. The Y6 films were spin coated at 1500 rpm for 70 s and then annealed for 20 minutes at 100 °C and left to cool until they were below 60 °C.

Cyclic voltammetry was conducted using a three-electrode setup, with the Y6 film as the working electrode, a Pt wire counter electrode, and an oxidized Ag wire reference electrode. For the oxidation measurements, a 0.1 M TBAPF6 solution in MeCN was used as the electrolyte. A PalmSens4 was used to run the electrochemical measurement. The electrochemical setup was placed inside a quartz cuvette, and the Y6 film was aligned in the beam path of a Bruker VERTEX 70v FT-IR spectrometer fitted with a Bruker DigiTect Mid-Infrared deuterated L-alanine doped triglycine sulphate (DLaTGS) detector with a KBr window. Scans were taken with a resolution of 4 $cm^{-1}$, aperture size of 3.5 mm, and scanner velocity of 5 kHz using the NIR source of the spectrometer.

**Raman and ATR-FTIR Experiments**
To perform Raman measurements, samples were prepared by drop-casting onto calcium fluoride substrates from the same 10 mg $mL^{-1}$ precursor solution as was used for the spin-coated samples. Drop-casting was necessary in order to achieve sufficient film thickness to yield a good SNR; Raman-grade calcium fluoride substrates, with a single characteristic peak at 321 $cm^{-1}$ were required in order to give a clean, unstructured background signal and hence reliable spectra. All spectra were measured on a Raman microscope (Horiba T64000) with 532 nm excitation. Wavenumber calibration was achieved by indexing the characteristic 520.7 $cm^{-1}$ Raman peak of a crystalline silicon wafer.

To perform attenuated total reflectance (ATR-FTIR) measurements, samples were prepared by spin-coating onto borosilicate float-glass substrates as described above. ATR-FTIR spectra were measured on an FTIR spectrometer (Bruker VERTEX 70v with ATR attachment). Additional details on the specific experimental adjustments are given in Section S13.

## Associated Content

**Supporting information.** Y6 crystal structure, Tuning of the long-range corrected functional, Reorganization energies and Huang-Rhys factors, Absorption spectrum, Fitting of Y6 single molecule spectrum, Frenkel-Holstein Hamiltonian for aggregate systems, Calculation of photoinduced electronic couplings, Calculation of charge transfer integrals, Calculation of excitation energies, Calculation of charge transfer excitations, Calculation of the character of the excited states, Convergence Y6 solid-state spectrum, Experimental spectrum spectrum of Y6 thin-film, Raman and ATR-FTIR, Transient absorption spectroscopy of Y6 in dilute media, Transient absorption spectroscopy of Y6 in BHJ blends, Additional Transient Absorption spectroscopy data on Y6 films, Spectroelectrochemistry measurements, Magnetic Resonance Spectroscopies.

## Acknowledgements

S.G. would like to acknowledge Jochen Blumberger, Lorenzo Cupellini, Nicholas J. Hestand, Raja Ghosh and Frank C. Spano for useful discussions at the beginning of the project. We thank Yiwen Wang for providing the donor:Y6 films used in this study. Andreas Sperlich and Vladimir Dyakonov are acknowledged for valuable discussions on PLDMR measurements throughout this work. We thank Ian Jacobs for assistance with the ATR-FTIR measurements. The work in Mons has been funded by the Belgian National Fund for Scientific Research (FRS-FNRS) within the Consortium des Équipements de Calcul Intensif (CÉCI), under Grant 2.5020.11, and by the Walloon Region (ZENOBE Tier-1 supercomputer), under grant 1117545. D.B. is a FNRS research director. Work in Grenoble has been supported by the French "Agence Nationale de la Recherche", project RAPTORS (ANR-21-CE24-0004-01) and used high-performance computing resources from GENCI-TGCC (grant no. 2022-A0130910016). S.G. thanks the support of ICSC‐Centro Nazionale di Ricerca in High Performance Computing, Big Data and Quantum Computing, funded by European Union‐NextGenerationEU‐PNRR, Missione 4 Componente 2 Investimento 1.4. A.J.G. thanks the Leverhulme Trust for an Early Career Fellowship (ECF-2022-445). D.J.C.S. and N.C.G. thank the EPSRC for support (grants EP/W017091/1 and EP/L01551X/2). J.G. acknowledges support by the Deutsche Forschungsgemeinschaft (DFG, German Research Foundation) within the Research Training School "Molecular biradicals: Structure, properties and reactivity" (GRK2112) and the Bavarian Ministry of the Environment and Consumer Protection, the Bavarian Network "Solar Technologies Go Hybrid".

## CRediT authorship contribution statement

**Samuele Giannini**: Investigation, Data Curation, Methodology, Writing – original draft; **Daniel J. C. Sowood**: Investigation, Data Curation; **Jesús Cerdá**: Methodology, Software; **Siebe Frederix**, Investigation, Data Curation; **Jeannine Grüne**: Investigation, Data Curation; **Giacomo Londi** Investigation, Data Curation; **Thomas Marsh**: Investigation; **Pratyush Ghosh**: Investigation; **Ivan Duchemin:** Methodology; **Neil C. Greenham**: Supervision; **Koen Vandewal:** Supervision, Writing

- Review & Editing; **Gabriele D'Avino:** Methodology, Supervision, Writing - Review & Editing; **Alexander J. Gillett**: Conceptualization, Investigation, Funding acquisition, Supervision, Writing – original draft; **David Beljonne** Conceptualization, Funding acquisition, Supervision, Writing - Review & Editing.

## Data Availability

Data will be made available on request at [Zenodo link to be added]

## Notes

The authors declare no competing financial interest.

## TOC

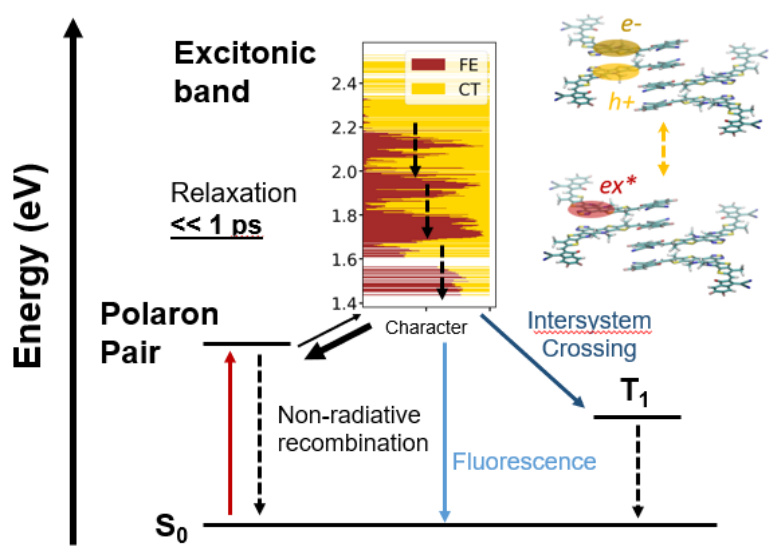

# Supplementary Information

### Section S1: Y6 crystal structure

The crystal structure of Y6 used in this work was taken from Ref.[1] The unit cell contains four molecules (Z = 4) and two distinct asymmetric units as shown in Fig. S1. To reduce the uncertainty of the experimental structure, bond lengths and angles of all the molecules in the unit cell were optimized using B3LYP/def2-SVP level of theory. While dihedral angles were kept unchanged to avoid altering the relative orientation of the molecules.

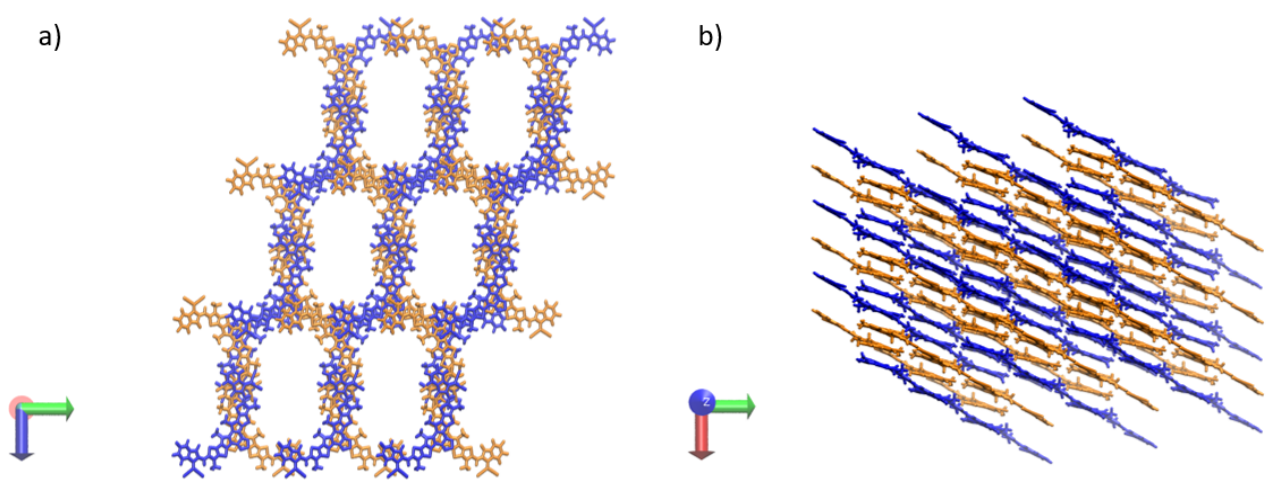

**Figure S1:** Y6 crystal structure. (a) Top view over the *yz* plane, (b) lateral view of the *xy* plane. Different colours are used for the two asymmetric units in unit cell. Y6 alkyl side chains are removed for clarity.

The unit cell was replicated to build a 2x2x2 supercell from which we extracted all unique dimers within a maximum centre of mass (COM) cut-off distance of 20 Å. Among these pairs, eight were direct contact pairs with a minimum atom distance shorter than 5 Å, while the others were non-contact pairs presenting other molecules or molecular fragments in between them (see Fig. S2). All the pairs were used to calculate various parameters entering the Hamiltonian (see main text).

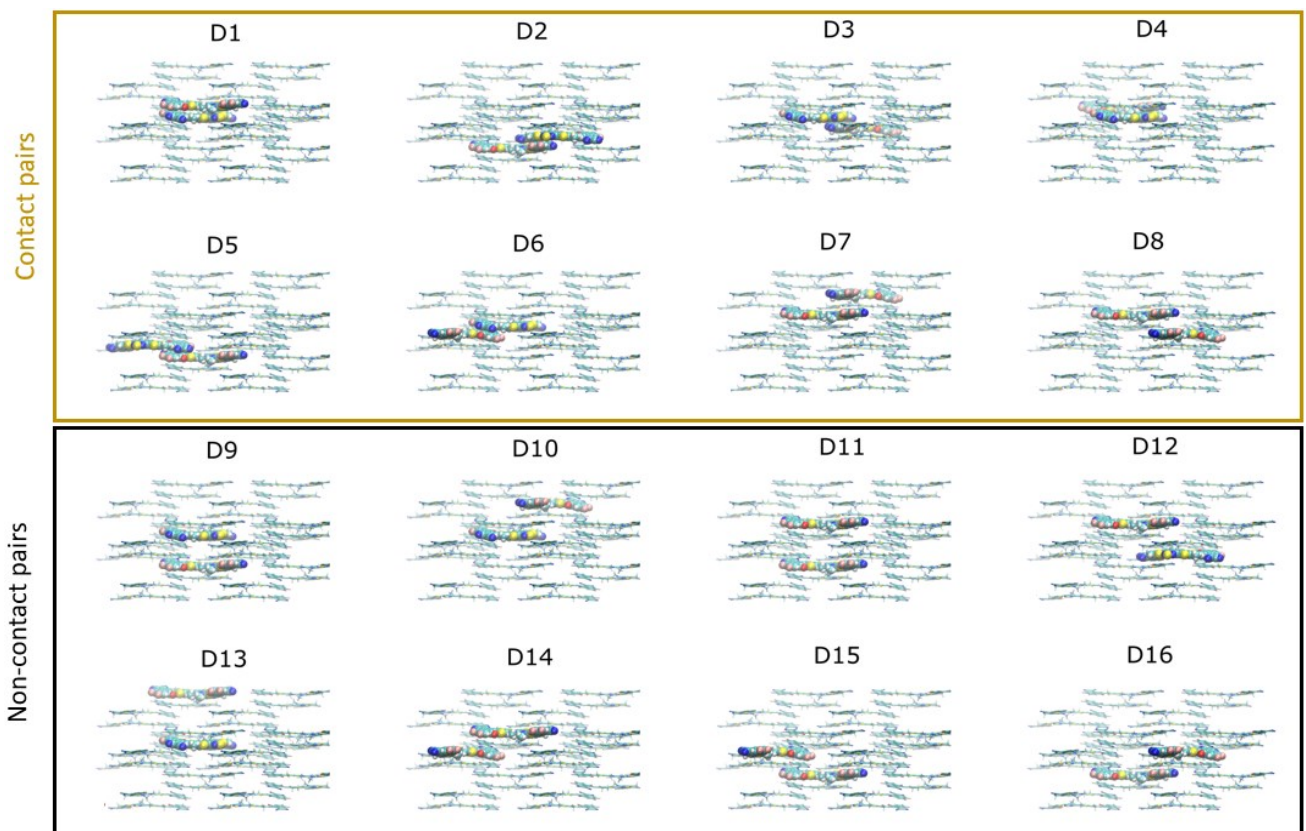

**Figure S2:** Y6 unique dimers with COM distance within 20 Å extracted from a 2x2x2 supercell.

## Section S2: Tuning of the long-range corrected functional

To improve the quality of the DFT functional for computing electronic couplings and reorganization energies of excited states of different character (see following sections), we used an optimally-tuned (OT) range-separated hybrid (RSH) functional. In such scheme,[2] the two-electron operator is divided into short- and long-range contributions, through the error function:

$$\frac{1}{r} = \frac{1 - (\alpha + \beta erf(\omega r))}{r} + \frac{\alpha + \beta erf(\omega r)}{r} \quad (S1)$$

where the parameters $\alpha$ and $\beta$ control the percentage of HF exchange at zero and infinite electron separation, respectively, and ω controls the switch-off distance between the short- and long-range parts. The semi-local, DFT-like exchange is included in the short-range domain, whereas the long-range part is described by the exact HF exchange. Here, the long-range potential behaves as $(\alpha + \beta)/r$, and the amount of long-range HF exchange is 100% at long inter-electronic distances. This is achieved by imposing $\alpha + \beta = 1$, which also ensures that the asymptotic behaviour of the charge-transfer energy is $E_{CT}(r) = IP - EA - 1/r$. The non-empirical tuning procedure adopted in this work is described elsewhere[3–5] and the OT ω value was found at 0.101 Bohr$^{-1}$ for the Y6 molecule by using the LC-ωhPBE functional and the 6-31G(d,p) basis set.[4]

## Section S3: Reorganization energies and Huang-Rhys factors

The reorganization energies associated with the excited (that is intended in this context as anionic, cationic and excitonic) states of a Y6 molecule were evaluated using the 4-point approach described, *e.g.*, in Ref.[6–8]:

$$\lambda^{4-\text{point}} = \lambda_{EX}^{\text{rel}} + \lambda_{N}^{\text{rel}} = [E_{EX}(\mathbf{R}_N) + E_N(\mathbf{R}_{EX})] - [E_{EX}(\mathbf{R}_{EX}) + E_N(\mathbf{R}_N)] \quad (S2)$$

where $\lambda_{EX}^{\text{rel}}$ and $\lambda_{N}^{\text{rel}}$ are the relaxation energies of the excited and neutral states, respectively. $E_{EX(N)}(\mathbf{R}_{N(EX)})$ is the energy of the excited (neutral) molecule in the optimized neutral (excited) state and $E_{EX(N)}(\mathbf{R}_{EX(N)})$ is the energy of excited (neutral) molecule in the optimized excited (neutral) minimum. These values were computed using our OT LC-ωhPBE/6-31G(d,p) and level of theory and are reported in Table S1.

A useful approach to find the parameters needed to build the Holstein Hamiltonian in Eq. 1 in the main text (*i.e.*, the Huang-Rhys factor and effective frequency) consists in decomposing the relaxation energies in terms of frequency-resolved normal modes. We refer to this approach as normal mode analysis (NMA). To this end, we used a displaced harmonic oscillator model where we assumed that both the excited and neutral states have the same potential energy's curvature and we projected each intramolecular normal mode on the vector describing the geometric changes between the neutral and excited states (*i.e.*, anionic, cationic or excitonic) to partition the relaxation energies into mode contributions.[9] In this way, we can write the total reorganization energy as twice the relaxation energy of the excited and neutral state (*i.e.*, $\lambda^{\text{NMA}} = 2\lambda^{\text{rel}}$). The relaxation energy ($\lambda^{\text{rel}}$) can be written as:

$$\lambda^{\text{rel}} \cong \sum_i \hbar\omega_i S_i \quad (S3)$$

where $S_i$ denotes the Huang-Rhys factor of a specific mode of frequency $\hbar\omega_i$. Table S1 shows that the total reorganization energy $\lambda^{\text{NMA}}$ agrees well with the $\lambda^{4-\text{point}}$ in Eq. S2 for all Y6 excitations. The normal mode frequencies and Huang-Rhys contributions to relaxation energies shown in Fig. S3 were computed with the MOMAP package[10] that follows a previously reported methodology.[11]

**Table S1:** Reorganization energies (in eV) from 4-point scheme and normal mode analysis.

|  | CATION | ANION | EXCITON |
|---|---|---|---|
| $\lambda^{4-\text{point}}$ | 0.243 | 0.237 | 0.157 |
| $\lambda^{\text{NMA}}$ | 0.259 | 0.254 | 0.172 |
| $\lambda^{\text{rel}}$ | 0.129 | 0.127 | 0.086 |
| $\lambda_{hf}^{\text{rel}}$ | 0.087 | 0.104 | 0.059 |
| $\lambda_{lf}^{\text{rel}}$ | 0.042 | 0.023 | 0.027 |
| $\hbar\omega_{\text{eff}}$ | 0.180 | 0.181 | 0.182 |
| $S_{\text{eff}}$ | 0.485 | 0.575 | 0.325 |
| $\sqrt{S_{\text{eff}}}$ | 0.696 | 0.758 | 0.570 |

In Fig. S3, it can be seen that the relaxation energies to create an excited state involve vibrational frequencies above 1000 cm$^{-1}$. Thus, as done by others,[12] we assumed that part of the total relaxation energy associated to the formation of an excited state is carried out by a single high-frequency mode with an effective frequency $\hbar\omega_{\text{eff}}$. This quantity was evaluated for each excitation by weighting the frequencies of each mode across the full spectrum in Fig. S3 by the corresponding Huang-Rhys factor $S_i$, namely $\hbar\omega_{\text{eff}} = \sum_i \hbar\omega_i S_i / \sum_i S_i$. The effective Huang-Rhys factor associated to the high-frequency relaxation energy ($\lambda_{hf}^{\text{rel}}$) becomes $S_{\text{eff}} = \lambda_{hf}^{\text{rel}}/\hbar\omega_{\text{eff}}$. The rest of the relaxation energy associated with low-frequency modes ($\lambda_{lf}^{\text{rel}}$) was assumed to be classical and, thus, to contribute to the homogeneous broadening of the spectrum. The quality of the effective Huang-Rhys factor obtained from this procedure was checked against the experiment since this quantity is related to the ratio between the intensity of the first vibronic band and the Franck-Condon (FC) transition by $S_{\text{eff}} = \frac{I_{1-0}}{I_{0-0}}$.[13] From the optical absorption in chloroform solution we obtained $S_{\text{eff}} = 0.33$ which agree well with our computed estimate. Moreover, our results are also consistent with what previously found in Ref.[14].

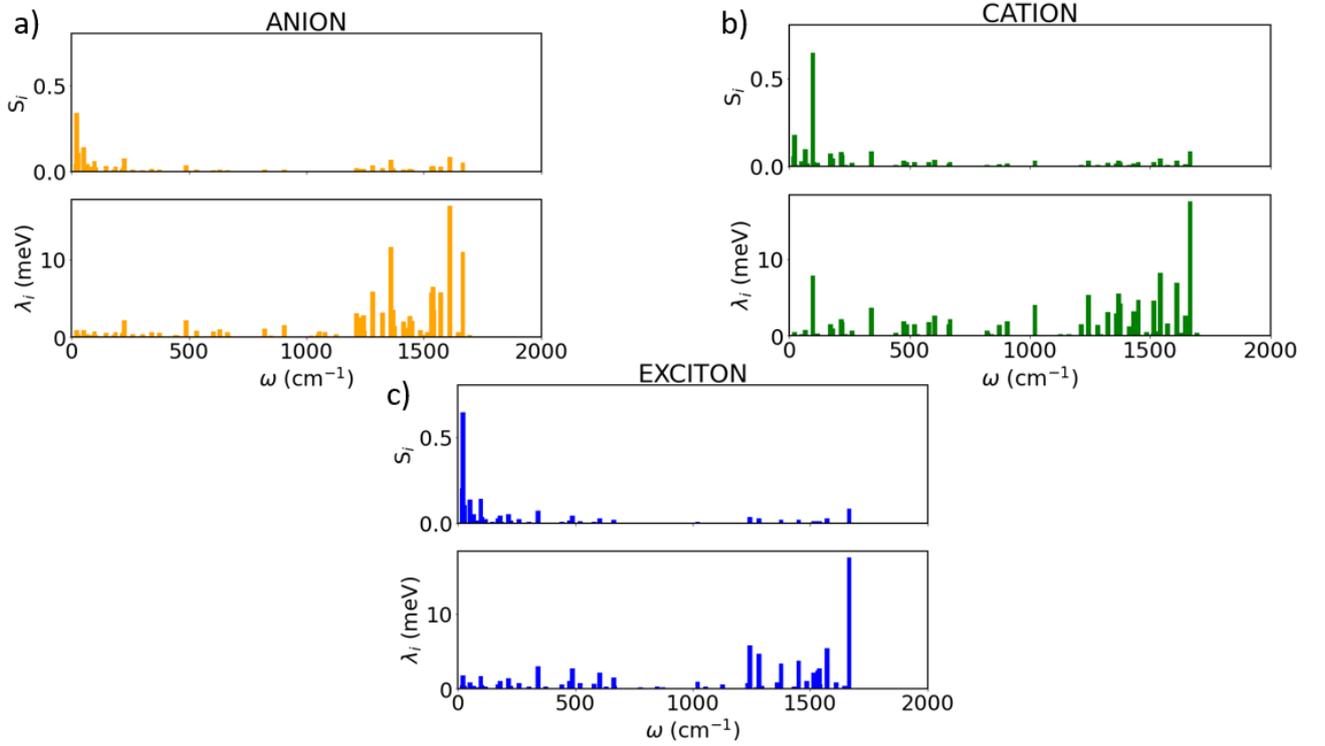

**Figure S3:** Huang-Rhys factors (top panels) and relaxation energies (bottom panels) when going from the neutral to the excited state geometry, namely **(a)** anionic, **(b)** cationic and **(c)** excitonic state.

## Section S4: Absorption spectrum

The absorption spectrum can be calculated by using:

$$A(E) = \frac{1}{|\mu|^2} \sum_j E_j |\langle \Psi^{(g)}|\hat{\mu}|\Psi^{(j)}\rangle|^2 W(E - E_j) \quad \text{(S4)}$$

where $\hat{\mu}$ is the transition dipole moment operator between the ground state wavefunction $\Psi^{(g)}$ and a given eigenstate $\Psi^{(j)}$ of the system (calculated using Eq. 8 in the main text), and $W$ is a line broadening function which in this work is taken to be a Gaussian function with a standard deviation σ (set to 60 meV). The line intensity is the oscillator strength calculated as the product of the square of the transition dipole moment and the transition energy ($E_j$).

## Section S5: Fitting of Y6 single molecule spectrum

For a single Y6 molecule the spectrum is computed using the Franck-Condon (FC) approximation which permits to write $\left|\langle \Psi^{(g)}|\hat{\mu}|\Psi^{(j)}\rangle\right|^2 = |\mu|^2 \exp(-S_{\text{eff}}) \frac{S_{\text{eff}}^{\tilde{v}}}{\tilde{v}!}$. When considering only one electronic transition ($S_0 \rightarrow S_1$), and one effective vibrational mode connected to the excitation the line shape of the absorption spectrum takes the form:

$$A(\omega) = |\mu|^2 \exp(-S_{\text{eff}}) \sum_{\tilde{v}=0}^{\tilde{v}_{max}} \frac{S_{\text{eff}}^{\tilde{v}}}{\tilde{v}!} \exp\left[-\frac{\left(E - \left(E_{S_0 \rightarrow S_1} + \tilde{v}\hbar\omega_{\text{eff}}\right)\right)^2}{2\sigma^2}\right] \quad (S5)$$

where $E_{S_0 \rightarrow S_1}$ is the $S_0 \rightarrow S_1$ electronic transition energy, $\tilde{v} = 0, 1, 2...$, denotes the number of vibrational energy levels of the excited states spaced by an effective vibrational frequency $\hbar\omega_{\text{eff}}$. $S_{\text{eff}}$ is the Huang Rhys factor of the electronic transition, which was calculated as previously described. To mimic the effect of the environment and related homogenous spectral broadening, a Gaussian line shape with constant standard deviation ($\sigma$) was assumed for all overtones. The comparison between theory and experiments resulting from this analysis is reported in Fig. S4.

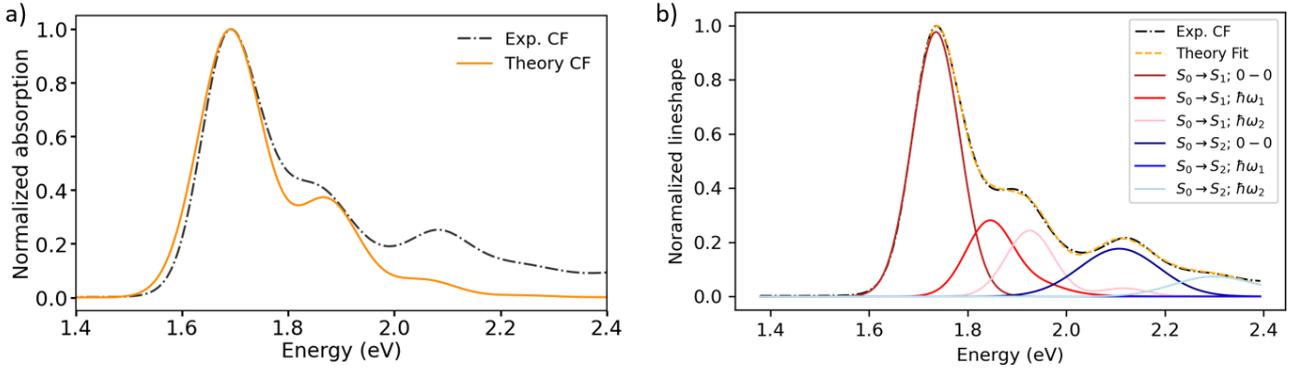

**Figure S4: (a)** Comparison between the experimental spectrum (gray line) of Y6 in chloroform and computed spectrum (orange line) with parameters taken from Table S1 for the exciton. The theoretical spectrum is shifted to match the energy of the 0-0 band in the experiment. **(b)** Fitting of the experimental spectrum using a multi-parameter FC fitting analysis.

For completeness, we note in passing that the simple FC model in Eq. S5 can be expanded to include multiple electronic transitions and multiple vibrational modes for each excitation, thereby systematically improving the agreement with experiment especially in the high-energy region of the spectrum (see Fig. S4b). For instance, it is well established[14,15] that the band at 2.08 eV is related to a second electronic transition ($S_0 \rightarrow S_2$), which carries its own vibronic progression, and we show this in Fig. S4b by decomposing the spectrum in its Gaussian components. Nevertheless, we note that since our objective is to simulate the spectral features of Y6 aggregate (refer to Methods), it is a reasonable approximation to include only a single electronic transition and its corresponding effective mode (Fig. S4a). This allows to keep the model computationally tractable when performing calculations on large super cells in the case of Y6 aggregate.

## Section S6: Frenkel-Holstein Hamiltonian for aggregate systems

The Hamiltonian is written as a sum of different terms describing different types of excitations on the chromophores and pairwise interactions between them (see Eq. 1 main text). $H_{\text{FE}}$ represents a standard

Frenkel exciton Hamiltonian with site energies: $E_{FE}^{(cry)} = E_{S_0 \to S_1}^{(sol)} + \Delta_{0-0}$ where $E_{S_0 \to S_1}^{(sol)}$ is the (adiabatic) energy of the $S_0 \to S_1$ electronic transition in solution and $\Delta_{0-0}$ is the solution-to-crystal red shift (STCS) that accounts for polarization and other electrostatic effects between molecules. The off-diagonal elements of this block are the excitonic couplings, $V_{kl}$, which describe the interactions between tightly a bound electron-hole pair sitting either on molecules $k$ or $l$ and allow for excitation energy transfer between the two. The second term $H_{CT}$ represents the charge transfer Hamiltonian (Eq. 3 main text) whose site energies, $E_{CT}(r_s)$, are the energies of an electron-hole pair with electron and hole at a given distance $r_s$, and the off-diagonal elements account for the electron and hole transfer integrals ($t_e$ and $t_h$, respectively) between molecules (Fig. S5). The mixing between FE and CT excitons is accounted for by the $H_{FE-CT}$ block (Eq. 4 main text). In this case, the coupling matrix elements permit exciton splitting through photoinduced electron transfer (PET) couplings ($D_e$) and photoinduced hole transfer (PHT) couplings ($D_h$) as shown in Fig. S5. Finally, FE as well as CT excitations are coupled to the nuclear degrees of freedom using appropriately computed by effective intramolecular modes tied to the formation of excitonic, cationic or anionic states on the different molecules (see Section S3).

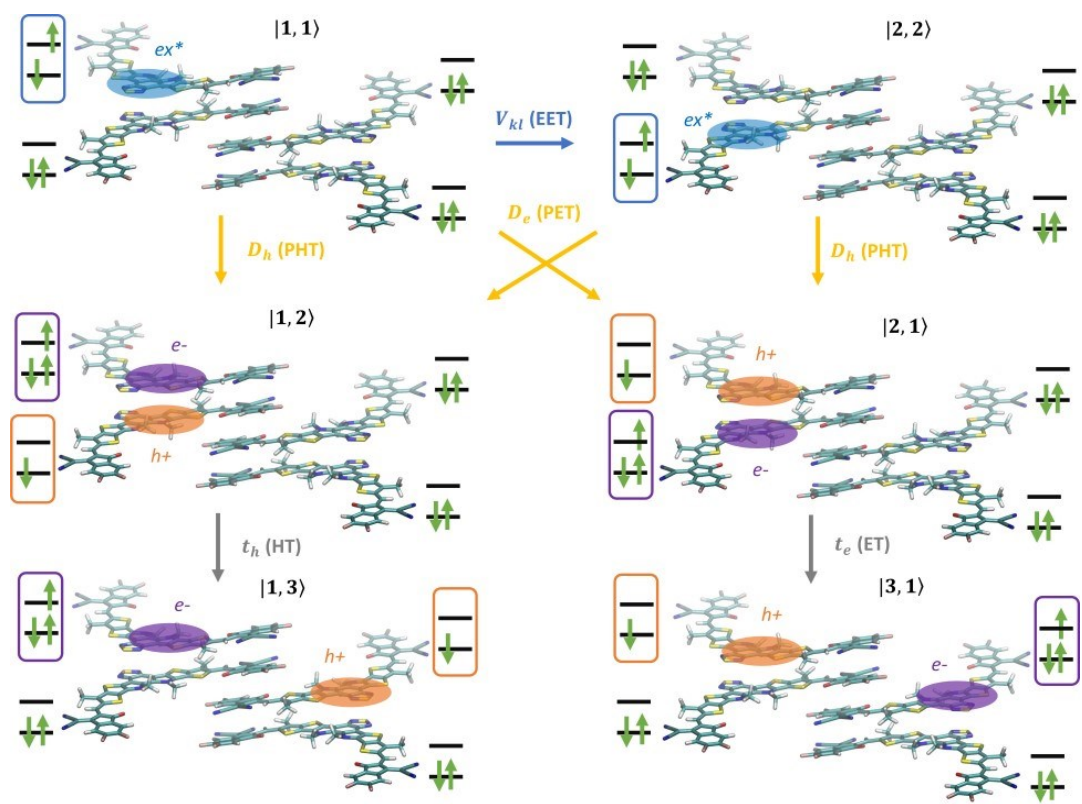

**Figure S5:** Schematic representation of the different processes and steps describe by the Frenkel-Holstein Hamiltonian in Eq. 1 in the main text. The coupling between localized Frenkel-type excitons ($V_{kl}$) is represented in blue (yielding excitation energy transfer (EET)). Photoinduced electron transfer (PET) and photoinduced hole transfer (PHT) processes responsible for exciton splitting and promoted by $D_e$ and $D_h$ electronic interactions are shown in yellow. Finally, electron and hole transfer processes with transfer integrals $t_e$ and $t_h$, respectively, are shown in grey.

## Section S7: Calculation of photoinduced electronic couplings

**Multi-state fragment excitation difference fragment charge difference approach**

The excitonic and PET and PHT couplings, namely $V_{kl}$, $D_e(s)$ and $D_h(s)$ respectively, are important parameters entering the Hamiltonian in Eq. 1. To compute these interactions, we employed the multi-state fragment excitation difference fragment charge difference approach (MS-FED-FCD) in combination with TDDFT (see computational details in Methods).[16–18] The MS-FED-FCD extends the capabilities of the previously developed diabatization schemes such as the 2-state fragment excitation difference (FED)[19] and fragment charge difference (FCD)[20,21] approaches. All these methods, which have been described in detail elsewhere[16–18,22], involved the partition of a given donor acceptor pair into two fragments and by using appropriate additional operators, the adiabatic Hamiltonian of the dimer (formed by two or more adiabatic states of the system) is transformed into a diabatic basis, which allows a direct evaluation of the couplings between FE states or CT states from the diabatic Hamiltonian matrix. However, when dealing with closely packed molecules, the adiabatic states of the interacting pair might be partially mixed with several states of different characters (*e.g.*, FE and CT excitons). This is where the MS-FED-FCD is particularly useful and goes beyond standard FED and FCD algorithms, which only allow to diabatize two states at a time (*i.e.*, two FEs or two CTs respectively). Specifically, MS-FED-FCD allows us to include several adiabatic excited states of the donor-acceptor system in the diabatization procedure to ensure a complete de-mixing between excitations of different nature and an optimal reconstruction of the localized FE and CT states, even in cases where a given adiabatic state is the combination of many diabatic states of both donor and acceptor (as it is the case here for Y6).

The MS-FED-FCD scheme permits to directly obtain $V_{kl}$, $D_e(s)$ and $D_h(s)$ couplings from a single TDDFT calculation on the investigated molecular pair as off-diagonal matrix elements of the Hamiltonian in diabatic state space. The TDDFT calculations required to obtain the above-mentioned couplings were done by using the Tamm-Dancoff approximation (TDA) and including 30 adiabatic excited states in the diabatization procedure. We employed our OT LC-ωhPBE/6-31G(d,p) level of theory consistently with the calculation of the relaxation energy.[4,23] We also compared these values with the standard CAM-B3LYP.

**Transition ESP charges**

Despite the usefulness of the MS-FED-FCD, its relatively high computational cost makes it impractical to compute long-range excitonic interactions between all pairs in a super cell of a large molecular aggregate (as the ones considered in this work). An alternative, approximate approach that has proven valuable to calculate Coulombic interactions in very good agreement with MS-FED-FCD in a number of organic crystals[22] relies on the interaction between transition charges obtained from the fitting of the electrostatic potential (TrESP). In this case the excitonic coupling is written as the Coulomb interaction of TrESP charges of donor and acceptor as:

$$V_{kl} = \frac{1}{\varepsilon_s} \sum_{A \in k} \sum_{B \in l} \frac{q_A^T q_B^T}{|\boldsymbol{r}_A - \boldsymbol{r}_B|} \tag{S6}$$

where the indices $A$ and $B$ run over the atoms of molecules $k$ and l, respectively, $q_A^T$, $q_B^T$ are the transition charges and $\boldsymbol{r}_A$, $\boldsymbol{r}_B$ are the positions of atoms $A$ and $B$, respectively. TrESP charges were obtained as proposed by Renger *et al.* in Ref[24], by fitting the electrostatic potential generated by the transition density. TrESP charges were calculated for the isolated molecule (employing our OT LC/6-31G(d,p) level of theory) and then used to compute all the long-range excitonic interactions in the Hamiltonian in Eq. 1 (even for pairs further away in space than those reported in Fig. S2). A screening factor $\varepsilon_s$, representing the dielectric constant, can be applied to include polarization effect due to the crystalline environment and herewe used the dielectric constant calculated using a microelectrostatic (ME) model[24,25] in the MESCal code as described in Methods.

**Table S2:** Y6 excitonic couplings for several Y6 pairs with two different levels of theory (i.e, OT LC-ωhPBE and CAM-B3LYP). The basis set is 6-31G(d,p).

| Pairs | Dist. [Ang] | OT LC-ωhPBE $V_{kl}^{MS-FED-FCD}$ [meV] | $V_{kl}$ [meV] | CAM-B3LYP $V_{kl}^{MS-FED-FCD}$ [meV] | $V_{kl}$ [meV] |
|---|---|---|---|---|---|
| D1 | 9.29 | -90.3 | -76.0 | -89.7 | -72.5 |
| D2 | 13.62 | 6.8 | 5.1 | 11.5 | 5.1 |
| D3 | 13.84 | -23.7 | -6.1 | -20.5 | -5.7 |
| D4 | 15.49 | -75.5 | -78.7 | -69.9 | -74.6 |
| D5 | 18.15 | 56.3 | 50.8 | 54.9 | 50.8 |
| D6 | 18.33 | 59.7 | 56.5 | 58.9 | 54.0 |
| D7 | 15.44 | -14.2 | -11.5 | -14.2 | -10.0 |
| D8 | 15.97 | -7.0 | -9.0 | -6.3 | -8.8 |
| D9 | 9.63 | -22.3 | -18.0 | -22.7 | -16.9 |
| D10 | 14.44 | -20.9 | -20.2 | -18.9 | -18.3 |
| D11 | 14.47 | 27.6 | 28.5 | 26.5 | 26.6 |
| D12 | 15.67 | -5.4 | -12.3 | -3.3 | -12.2 |
| D13 | 17.98 | -8.3 | -7.3 | -8.2 | -6.8 |
| D14 | 18.03 | -26.1 | -28.4 | -24.3 | -26.8 |
| D15 | 19.25 | -30.1 | -30.0 | -28.6 | -28.3 |
| D16 | 19.94 | 6.9 | 6.5 | 6.2 | 5.7 |
| MUE[1] | | | 4.19 | | 4.94 |

[1] Mean unsigned error defined as in Ref. [27] Note that as described in the text, the sign is consistently given for all the interactions and among different pairs, but is not adjusted to show H- vs J-like interactions (see Fig. S6).

In general the excitonic coupling sign between two molecules of a given dimer is arbitrary as it depends on the phase of the interacting molecular transition densities. Though, it is important to ensure a consistent relative phase and coupling sign when constructing the exciton Hamiltonian (Eq. 1) of a crystal with translational symmetry (as this has an important impact on the electronic and optical properties[28,29]). This requirement is straightforwardly fulfilled when using atomic TrESP charges to represent transition densities and to calculate excitonic couplings, $V_{kl}$. In fact, with this method the phase of the molecular transition density is determined by the atom ordering, which is kept consistent by construction for all the molecules in the supercell. Moreover, the TrESP approach allows in principle to univocally distinguish positive (which we called H-like) vs negative (indicated with J-like) interactions computing a signed transition-dipole-corrected coupling $\tilde{V}_{kl}$ as

$$\tilde{V}_{kl} = \text{sgn}(\boldsymbol{d}_k \cdot \boldsymbol{d}_l) V_{kl} \tag{S7}$$

where the term in between parenthesis is the scalar product of the transition dipoles obtained using atomic TrESP charges as $\boldsymbol{d} = \sum_I q_I^T \boldsymbol{r}_I$ ($I$ runs over the atomic positions of the molecule and $q_I^T$ are the atomic TrESP charges). See Fig. 2 in the main text and Fig. S6 for a visual representation of H- vs J-like interactions.

In Table S2, we report the comparison of the excitonic couplings evaluated with two different level of theory (i.e, OT LC-ωhPBE and CAM-B3LYP) and related TrESP couplings. Both levels of theory give comparable couplings, underscoring the fact that these excitonic interactions are primarily influenced by the shape of the interacting transition densities, but not as much by the energetics of the interacting states. Note that the sign in Table S2 is consistently given for all the interactions and among different pairs. However, one needs to be careful as a negative excitonic coupling does not necessarily mean J-like interaction in this Table as the coupling sign depends also on the relative orientation of the interacting transition dipoles. One would need to use Eq. S7 to univocally distinguish H- vs J-like interactions (see also Fig. S6 where the signed coupling $\tilde{V}_{kl}$ is reported).

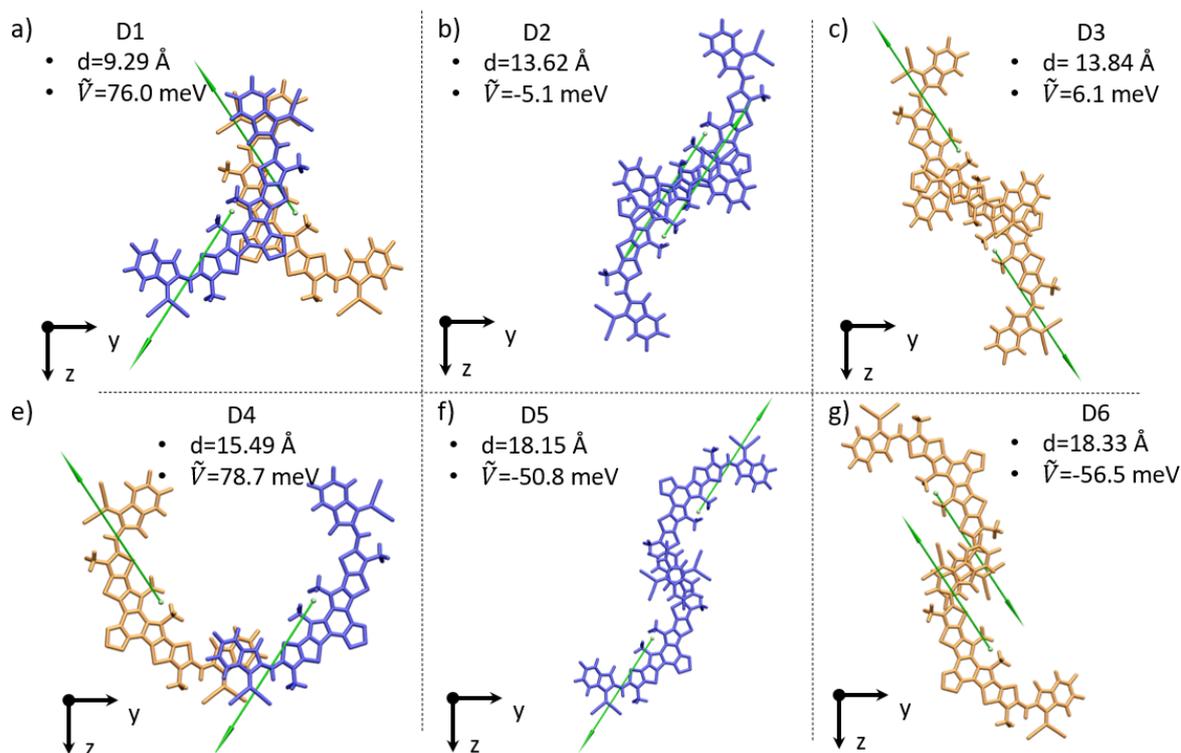

**Figure S6:** Y6 contact pairs and related interacting transition dipoles (scaled by an arbitrary amount for visualization purpose). The signed coupling is given according to Eq. S7. Thus, in this figure positive coupling sign represent H-like interacting molecules, while negative couplings J-like interacting molecules. Different colours are used for the two different asymmetric units in the unit cell.

## Section S8: Calculation of charge transfer integrals

The charge transfer integrals (*i.e.*, electron and hole transfer) between a pair of molecules are given by the Hamiltonian matrix elements:

$$t_e = \langle \phi_k^{\text{LUMO}} | \hat{H}_{ks} | \phi_l^{\text{LUMO}} \rangle \quad (S8)$$

and

$$t_h = -\langle \phi_k^{\text{HOMO}} | \hat{H}_{ks} | \phi_l^{\text{HOMO}} \rangle \quad (S9)$$

where $\phi_{k(l)}^{\text{HOMO(LUMO)}}$ represents the HOMO (LUMO) wavefunctions on a given pair of interacting molecules, and $\hat{H}_{ks}$ is the Kohn-Sham Hamiltonian of the system. Hole and electron transfer couplings in Eq. S8 and S9 were obtained using the projection operator-based diabatization (POD) approach, which was developed in Ref.[30–32] and extensively validated against high-level ab initio methods for a wide range of organic molecules containing heteroatoms.[31] The POD approach is a diabatization scheme similar in spirit to the MS-FED-FCD. POD uses the self-consistent DFT Kohn-Sham orbitals of the donor-acceptor pair under investigation. The corresponding Hamiltonian is represented in an orthogonalized atomic orbital basis and partitioned into donor and acceptor blocks (related to the two interacting fragments). After separately diagonalizing each block and transforming the off-diagonal blocks using unitary transformations, the resulting off-diagonal matrix elements are identified as electronic couplings between the block-diagonal donor and acceptor states. Hole transfer integrals are obtained by taking the Hamiltonian matrix element between POD localized HOMO orbitals, while the electron transfer couplings are obtained by taking the interaction between the LUMOs. The DFT calculations to get the charge transfer integrals were carried out as done in previous works[31,33,34] with the PBE functional,[35] DZVP basis set and GTH pseudopotentials on the neutral dimer of the interacting molecules (CP2K program[36] package was used for these calculations). As recommended in Ref.[31], POD couplings were scaled by a factor of 1.325 for the best agreement with high-quality ab-initio coupling estimates. These data are reported in Table S3 for the Y6 with non-zero interactions.

Importantly, a consistent relative sign of the couplings for different pairs was ensured by imposing the same phase of the HOMO and LUMO orbitals on all the molecules in the crystal (by exploiting their consistent atomic ordering analogously to what already mentioned for the TrESP couplings). In such a way, the coupling signs, which are determined by the phase of the interacting orbitals, are consistent for all the pairs in the model and with the excitonic coupling calculated using atomic TrESP charges. Finally, the same phase convention used for $t_e$ and $t_h$ was used for the photoinduced electron and hole transfer couplings ($D_e$ and $D_h$) throughout the model making the off-diagonal elements of the full Hamiltonian in Eq. 1 of the main text fully consistent among each other. This being an extremely important issue for obtaining reliable calculations.

**Table S3** Excitonic interactions, photoinduced electron and hole transfer couplings and electron and hole transfer integrals for the closest-contact Y6 dimers, D1 to D8, with given center of mass distance.

| Dimers | Dist. [Ang] | $V_{kl}$ [meV] | $D_e$ [meV] | $D_h$ [meV] | $t_e$ [meV] | $t_h$ [meV] |
|---|---|---|---|---|---|---|
| D1 | 9.29 | -76.0 | 72.0 | 55.7 | 66.1 | 49.6 |
| D2 | 13.62 | 5.1 | -55.4 | -45.5 | -79.8 | -7.4 |
| D3 | 13.84 | -6.1 | -15.0 | -27.3 | -27.0 | -11.3 |
| D4 | 15.49 | -78.7 | 53.1 | -10.9 | 71.4 | -6.7 |
| D5 | 18.15 | 50.8 | -68.9 | 24.3 | -66.1 | 20.0 |
| D6 | 18.33 | 56.5 | -47.9 | 33.5 | -42.1 | 19.2 |
| D7 | 15.44 | -11.5 | 0.0 | 0.0 | 0.0 | 0.0 |
| D8 | 15.97 | -9.0 | 0.0 | 0.0 | 0.0 | 0.0 |

The sign here is consistently given for all the different kind of interactions and among different pairs (*i.e.*, the sign is consistent with the phase of the basis set used). However, one needs to be careful as a negative excitonic coupling does not necessarily mean J-like interaction in this context as one needs to account for the fact that the coupling sign depends also on the relative orientation of the interacting transition dipoles as previously discussed (see Fig. S6 for the assignment of the sign using Eq. S7).

## Section S9: Calculation of excitation energies

Embedded many-body GW and Bethe Salpeter equation (BSE) calculations have been performed as described in the Method section of the main text for the single molecules in the crystal and reported in Table S4 and S5.

**Table S4:** Quasiparticle energy levels from GW/MM calculations on the Y6 crystal, considering one molecule as QM region. The table reports the different contributions to the crystal energy levels as defined in Ref.[37], namely the isolated molecule energy (gas) and the electrostatic (or crystal field, $D_E$) and polarization ($D_P$) contributions from the environment. The difference in energy levels between the two asymmetry units is about 0.1 eV.

| molecule | level | gas | $D_E$ | $D_P$ | crystal |
|---|---|---|---|---|---|
| Y6-1 | HOMO | -6.563 | -1.004 | 0.699 | -6.868 |
| | gap | 4.214 | -0.034 | -1.639 | 2.541 |
| | LUMO | -2.349 | -1.038 | -0.940 | -4.327 |
| Y6-2 | HOMO | -6.465 | -0.999 | 0.686 | -6.778 |
| | gap | 4.095 | -0.042 | -1.625 | 2.427 |
| | LUMO | -2.370 | -1.042 | -0.939 | -4.351 |

**Table S5:** Lowest-energy singlet excitations from BSE/GW/MM calculations on the Y6 crystal, considering one molecule as QM region. The Table reports excitation energies in eV and the corresponding oscillator strengths. The energy difference between the bright lowest-energy excitons of the two asymmetry units is about 0.1 eV, consistent with gap values.

| Y6-1 | | Y6-2 | |
|---|---|---|---|
| Energy | $f_{osc}$ | Energy | $f_{osc}$ |
| 1.67 | 2.11 | 1.59 | 2.29 |
| 2.08 | 0.46 | 2.02 | 0.38 |
| 2.39 | 0.04 | 2.39 | 0.04 |
| 2.42 | 0.10 | 2.39 | 0.00 |
| 2.54 | 0.35 | 2.46 | 0.08 |
| 2.58 | 0.08 | 2.50 | 0.40 |

The adiabatic energies of the excited states for some of the Y6 dimers extracted as described in Section S1 were calculated using high-quality many-body GW/BSE calculations and a polarizable molecular embedding (MM) strategy based on the microelectrostatic approach as described in Methods.[25,26,37]

**Table S6:** S1 exciton energies from embedded BSE calculations in PCM (at the same geometry and for different values of the optical dielectric constant) and in the gas phase (for different geometries). These results permit to assess the different contributions to the STCS, $\Delta_{0-0} = -160$ meV. The higher dielectric constant of the Y6 crystal ($\varepsilon_r^{(opt)} \approx 3$ as compared to 2.1 for CF) accounts for about -80 meV to the STCS. The effect of the crystal electrostatic field on the gap ($D_E$ in Table S4) contribute for about -38 meV (average over the two molecules). The effect of molecular geometries is, on average, comparable to that of the crystal field.

| Geometry | $\varepsilon_r^{(opt)}$ | $E'_{S_0 \to S_1}$ (eV) | $E'_{S_0 \to S_1} - E'^{,(ref)}_{S_0 \to S_1}$ (meV) | |
|---|---|---|---|---|
| Solution (PCM-CF) | 2.1(CF) | 1.790 | 0 | reference |
| Solution (PCM-CF) | 3.0 | 1.710 | -80 | |
| Solution (PCM-CF) | 4.0 | 1.675 | -115 | |
| Solution (PCM-CF) | 5.0 | 1.652 | -138 | |
| Solution (PCM-CF) | 1 (gas) | 2.043 | 0 | reference |
| Crystal (Y6A-1) | 1 (gas) | 2.046 | 3 | |
| Crystal (Y6A-2) | 1 (gas) | 1.982 | -61 | |

**Table S7:** GW/BSE/MM vertical energies in dimers on dimers.

| | D1 | | D2 | | D3 | | D4 | |
|---|---|---|---|---|---|---|---|---|
| State | Energy (eV) | $f_{osc}$ | Energy (eV) | $f_{osc}$ | Energy (eV) | $f_{osc}$ | Energy (eV) | $f_{osc}$ |
| S1 | 1.52 | 0.62 | 1.67 | 0.57 | 1.62 | 0.16 | 1.60 | 1.76 |
| S2 | 1.68 | 2.99 | 1.72 | 4.05 | 1.66 | 4.70 | 1.68 | 2.70 |
| S3 | 1.85 | 0.24 | 1.92 | 0.03 | 1.84 | 0.02 | 1.95 | 0.31 |
| S4 | 1.89 | 0.26 | 1.99 | 0.01 | 1.89 | 0.01 | 2.02 | 0.26 |
| S5 | 2.01 | 0.88 | 2.08 | 0.07 | 2.05 | 0.00 | 2.10 | 0.28 |
| S6 | 2.11 | 0.05 | 2.15 | 0.56 | 2.13 | 0.53 | 2.19 | 0.17 |
| | D5 | | D6 | | D9 | | D10 | |
| State | Energy (eV) | $f_{osc}$ | Energy (eV) | $f_{osc}$ | Energy (eV) | $f_{osc}$ | Energy (eV) | $f_{osc}$ |
| S1 | 1.63 | 4.06 | 1.55 | 4.50 | 1.59 | 1.85 | 1.59 | 0.19 |
| S2 | 1.72 | 0.08 | 1.64 | 0.01 | 1.66 | 2.36 | 1.60 | 4.41 |
| S3 | 2.02 | 0.45 | 1.97 | 0.38 | 2.02 | 0.38 | 2.01 | 0.38 |
| S4 | 2.02 | 0.20 | 2.00 | 0.23 | 2.07 | 0.47 | 2.03 | 0.38 |
| S5 | 2.10 | 0.01 | 2.04 | 0.13 | 2.07 | 0.00 | 2.09 | 0.00 |
| S6 | 2.12 | 0.33 | 2.08 | 0.13 | 2.11 | 0.00 | 2.14 | 0.00 |

## Section S10: Calculation of charge transfer excitations

To complete the Hamiltonian in Eq. 1 the energy of the diabatic FE and CT states ($E_{\text{FE}}^{(\text{cry})}, E_{\text{CT}}(r_s)$) are needed. TDDFT is known to be problematic in dealing with CT excitations and we found that, in contrast to the electronic coupling interactions, which are rather independent on the choice of the functional (see Table S2), the quantitative position of intermolecular CT-like adiabatic states within the excited state manifold of Y6 dimers changes considerably by varying the TDDFT level of theory. Moreover, the dielectric response of the environment additionally impacts the energy and character of the states and should be taken into account. To overcome these issues, we used BSE calculations with polarizable MM embedding performed on some of the pairs extracted from the crystal. We considered the closest contact pairs with non-negligible FE-CT couplings, and we represented their corresponding GW/BSE/MM spectra obtained by using the calculated adiabatic energies and oscillator strengths (reported in Table S7) for the lowest energy states in Fig. S7. We then set up a minimal tight-binding Hamiltonian formed by two FE states and two CT states and their corresponding couplings (taken from Table S7). The (diabatic) site energies of the corresponding states of different characters were adjusted and the model Hamiltonian iteratively diagonalized to reach the best possible agreement with the GW/BSE/MM spectra. This procedure, although based on a minimal model, allows us to obtain $E_{S_0 \to S_1}^{\prime,(\text{cry})}, E_{\text{CT}}^\prime(r_s)$ at the quality of the GW/BSE including polarizable electrostatic embedding, thereby going beyond TDDFT. We note in passing, that the prime in $E_{\text{FE}}^{\prime,(\text{cry})}, E_{\text{CT}}^\prime(r_s)$ is used to indicate that these are vertical excitation energies to which one needs to subtract the relaxation energy of the corresponding excitation in Table S1 to obtain the 0-0 energies $E_{\text{FE}}^{(\text{cry})}, E_{\text{CT}}(r_s)$ in the Hamiltonian.

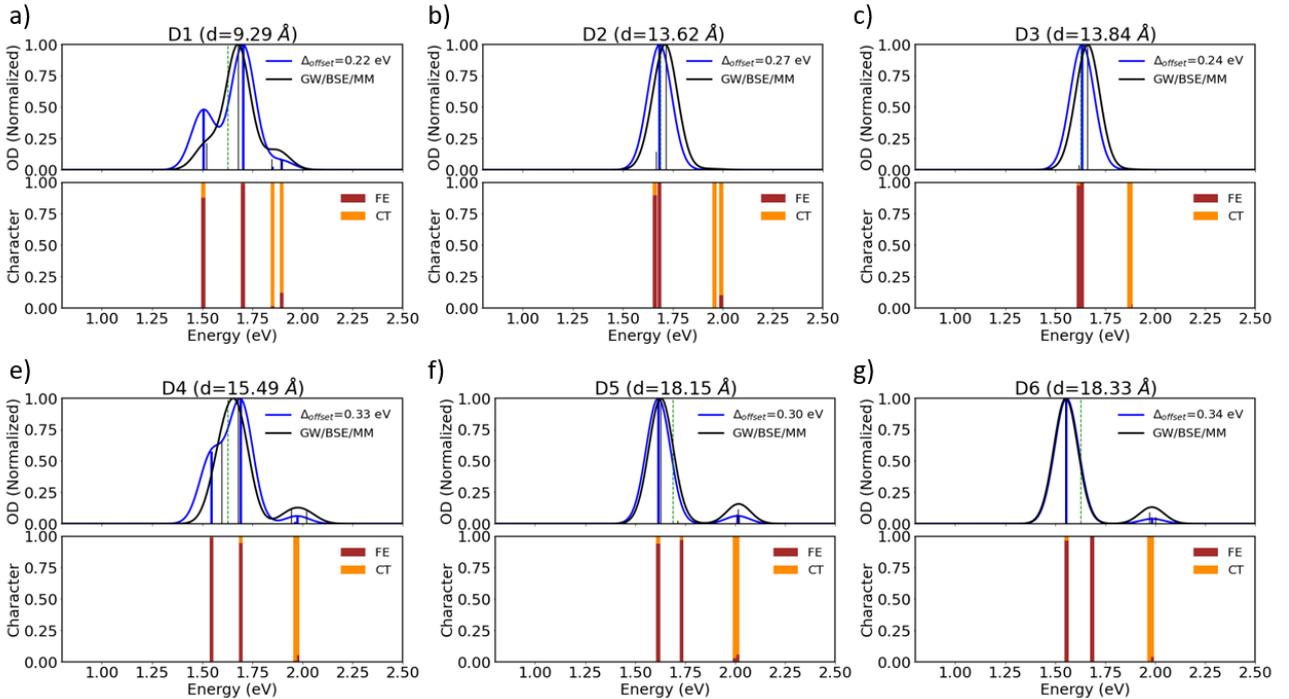

**Figure S7:** Y6 contact pairs and related absorption spectra calculated as described in the text using GW/BSE/MM data from Table S7 (solid black line) and a reconstructed tight binding Hamiltonian (solid blue line) as described in the text.

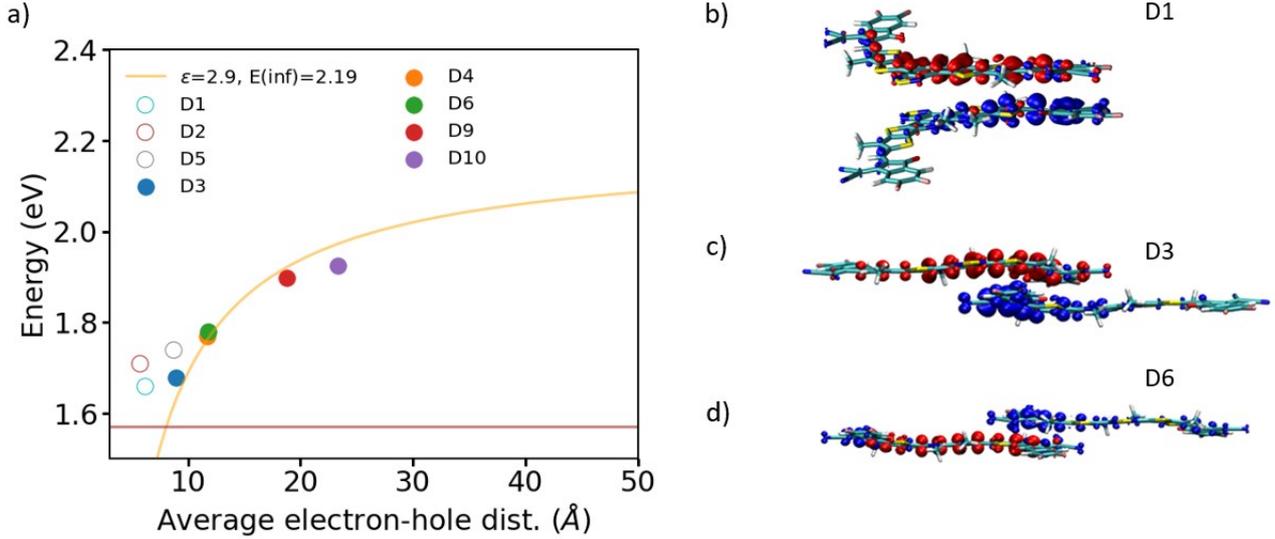

**Figure S8: (a)** Energy of charge transfer states, $E_{\text{CT}}(r_s)$, for some of the unique pairs extracted from the crystal as a function of electron-hole separation (calculated from the centroids of the attachment and detachment densities). The orange line represents a Coulomb fitting (with a fixed screening constant ε = 2.9) to the filled coloured circles. $E_{\text{CT}}(r_s)$ was explicitly calculated using GW/BSE/MM with the addition of the relaxation energies of cation and anion (Table S1). Horizontal brown line marks the Frenkel exciton energy ($E_{\text{FE}}^{(\text{cry})}$) used for all the molecules in the crystal and calculated from GW/BSE/MM. Panels **(b), (c), (d)** depict electron and hole particle wavefunctions of three different pairs at various distances.

The calculated values are reported in Fig. 5 with coloured circles as a function of the distance between the electron and hole centroids of the interacting molecules. The electron and hole particle wavefunctions for the lowest energy CT states of few selected pairs are also shown in Fig. S8b,c,d. Because these quantum-chemical calculations are highly costly, we opted for an explicit calculation of the CT energy of the closest contact pairs (for which the FE-CT interactions are sizable) and a few extra pairs only. The $E_{\text{CT}}(r_s)$ for all the other dimers in the supercell was estimated using an effective (screened) Coulomb barrier which is expected to work well to estimate the interaction between electron and hole at longer intermolecular distances. The filled circles data points were fitted with the Coulomb expression as:

$$E_{\text{CT}}(r_s) = E_{\text{CT}}(\infty) - \frac{e^2}{4\pi\varepsilon_0}\frac{1}{\varepsilon r_s} \quad (S10)$$

The dielectric constant in the fitting expression was fixed to the same value used to screen the excitonic interactions (calculated with a classical microelectrostatic model as described in Methods) and the energy at infinite electron-hole separation ($E_{\text{CT}}(\infty)$) was used as the only fitting parameter. From this analysis we estimated an exciton binding energy, $E_b$, of ~0.5 eV and we observe that most of the dimers nicely follow the expected trend. The empty circles in Fig. S8 represent the closest contact pairs deviating from such a Coulomb behaviour. This is not surprising considering the large size of the molecules, compared to the intermolecular distance, and the possible quantum mixing of elementary electron-hole excitations within the BSE framework.

## Section S11: Calculation of the character of the excited states

The character of the *j*-th eigenstate ($\xi$) of the Hamiltonian in Eq. 1 can be found by taking the square of the one and two particle coefficients as:

$$\xi^{(j)}_{FE(1p)} = \sum_{k} \sum_{\widetilde{v}=0}^{v_{max}} |c^{(j)}_{k,\widetilde{v}}|^2 \tag{S11}$$

$$\xi^{(j)}_{FE(2p)} = \sum_{k} \sum_{l \neq k} \sum_{\widetilde{v}=0}^{v_{max}-1} \sum_{v=1}^{v_{max}-\widetilde{v}} |c^{(j)}_{k,\widetilde{v},l,v}|^2 \tag{S12}$$

$$\xi^{(j)}_{CT} = \sum_{k} \sum_{l \neq k} \sum_{\widetilde{v}^+=0}^{v_{max}} \sum_{\widetilde{v}^-=0}^{v_{max}} |d^{(j)}_{k,\widetilde{v}^-,l,\widetilde{v}^+}|^2 \tag{S13}$$

## Section S12: Convergence of Y6 solid-state spectrum

The convergence of the Y6 aggregate spectrum in terms of supercell size is shown in Fig. S9a. Since our model includes periodic boundary conditions (introduced with a minimum image convention), the spectrum of the aggregate converges reasonably quickly with cell size. The amount of CT states in the Hamiltonian in Eq. 1 scales with $N^2$ where N is the number of molecules in the sample. To reduce the amount of CT states and keep the Hamiltonian diagonalization computational feasible we introduced a COM cut-off radius in the CT-basis beyond which the two molecules are considered uncoupled. We remind that charge transfer integrals decrease exponentially with the distance between molecules, so this alteration is not expected to sizably affect the shape of the spectrum as long as the cut-off radius is sufficiently large to incorporate all those interactions that show non-negligible $t_e$, $t_h$ and $D_e$, $D_h$ interactions. This is demonstrated in Fig. S9b where we depict the impact of the cut-off radius on the shape of the Y6 spectrum. We also note that no cut-off radius has been applied to the interaction between FE states, as these interactions are long-range and should be all taken into account. In Fig. S9c we show how the forcing photoinduced couplings to be the same of charge transfer interactions influence the shape of the spectrum and in Fig. S9d the convergence with the number of vibrational quanta considered in the basis set.

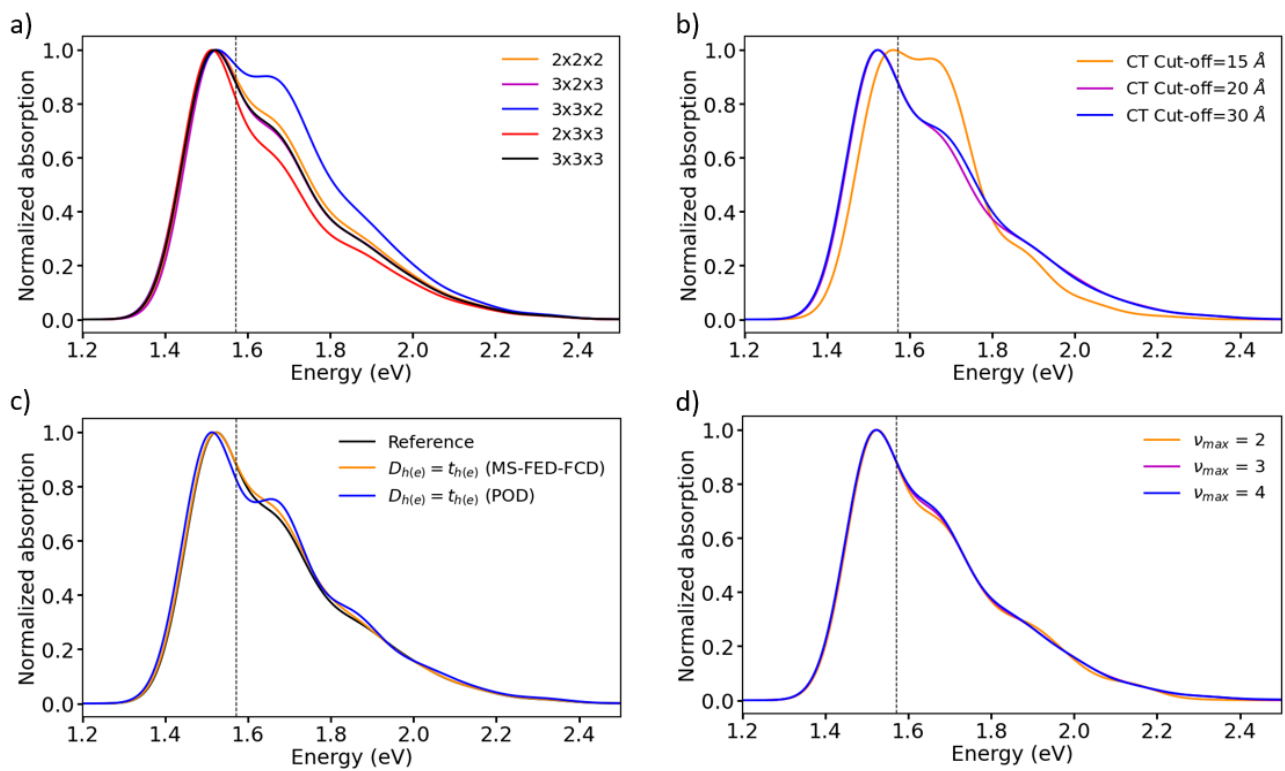

**Figure S9: (a)** Convergence of the Y6 aggregate spectrum with respect to supercell size and **(b)** CT-basis cut-off radius introduced as discussed in the text. Panel **(c)** shows the comparison of the Y6 aggregate spectrum when $D_{e(h)}$ and $t_{e(h)}$ are calculated as described in Section S7 (orange line) vs the case in which $D_{e(h)}$ and $t_{e(h)}$ are assumed to be the same and computed with either MS-FED-FCD or POD diabatization schemes. **(d)** Convergence with the number of vibrational quanta considered in the Holstein Hamiltonian.

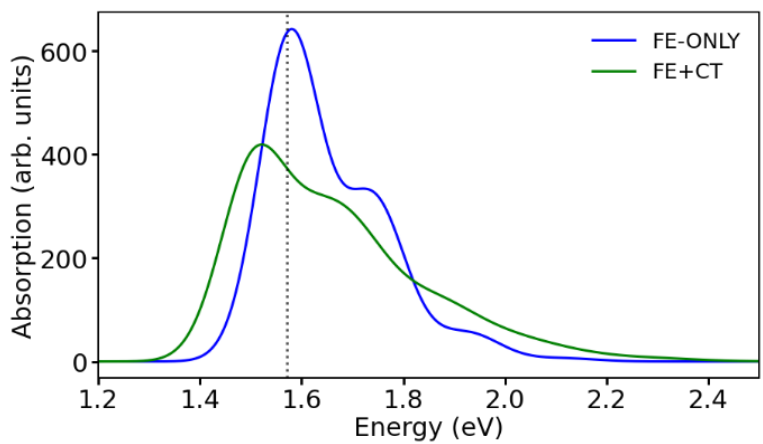

**Figure S10:** Computed unnormalized Y6 aggregate spectra without (blue line) and with (green line) CT states included in the Hamiltonian.

### Section S13: Experimental spectrum of Y6 thin-film

To measure the solution absorption spectra, a dilute solution of Y6 in chloroform (CF) and toluene was made in a high-performance quartz cuvette (Hellma Analytics). The absorption spectrum was determined via a transmission measurement using an UV-Vis-NIR spectrometer (Agilent Technologies Cary 5000). The concentration of Y6 was determined by diluting the solution so that the signal was not saturated as well as invariant under consecutive dilutions. The results are shown in Fig.

1c of the main text. Besides the Y6 thin-film obtained from CF which recent works showed[38,39] to yield a lamellar stacks face-on the substrates and to have a good crystalline morphology, for completeness we also prepared Y6 film from Chlorobenzene (CB). The film spectra were obtained using a combination of reflection-transmission measurements using an integrating sphere, and reverse Transfer Matrix Modeling (rTMM) to obtain the complex index of refraction of the films. Two/three separate films with different thicknesses were fabricated for each solvent. This was done by spin casting the material on borosilicate glass substrates at 2000 RPM. All films were annealed at 120°C. The different thicknesses were obtained by dissolving the Y6 in solutions with different concentrations. The processing parameters are listed in Table S8. The thicknesses in this table were measured using a profilometer (Bruker DekTakXT).

**Table S8:** Processing conditions and thicknesses of the series of Y6 films.

| Solvent | Concentration [mg/mL] | Thickness [nm] |
|---|---|---|
| Chloroform | 10.0 | 67 |
| | 7.5 | 53 |
| | 5.0 | 37 |
| Chlorobenzene | 10.0 | 59 |
| | 7.5 | 35 |

The transmission and reflection spectra of all films were then measured using the same UV-Vis-NIR spectrometer, now using an integrating sphere. Usually, one then obtains the absorption coefficient of the film by applying the Lambert Beer law to the reflection and transmission data. The problem is that this approach does not take into account thin film interference effects, meaning that the obtained absorption spectra are thickness dependent as can be seen in Fig. S11b. Kerremans et. al. developed a technique that allows for accurate determination of the complex index of refraction of thin film samples.[40] Here, the complex index of refraction is obtained by iterative adjustment so that when it is used for the TMM calculation, the experimental transmission spectrum of the film is optimally reproduced. This requires transmission (and optionally the reflection) data of the material with two different thicknesses. The absorption coefficient $A$ of the films were determined from the complex part of the complex index of refraction (the attenuation coefficient $k$)[40]:

$$A(\lambda) = \frac{4\pi k(\lambda)}{\lambda} \tag{S12}$$

The (real) index of refraction $n(\lambda)$ and attenuation coefficient $k(\lambda)$ shown in Fig. S11c were obtained via the program *NK_Finder* developed by Kerremans et. al. and distributed freely via ref.[40]

The 35 nm Chlorobenzene (CB) processed Y6 film showed an abnormally high normalized absorption intensity of approximately 0.1 below the gap as seen in Fig. S11b. The CB films were therefore excluded from the rTMM analysis. The rTMM analysis was performed on combinations of the CF film data resulting in three separate spectra as seen in Fig. S11d where they are compared to spectra obtained via the Lambert Beer law. Here, the interference corrected spectra show a higher 0-1 and 0-2 intensity compared to the spectra obtained from the Lambert Beer law, highlighting the importance of dealing with the interference effects. Additionally, they are significantly less dependent on the film thickness, and we speculate that the remaining difference is due to morphology and crystallinity of the film itself being slightly dependent on the film thickness as also discussed by Kerremans et. al.[40]

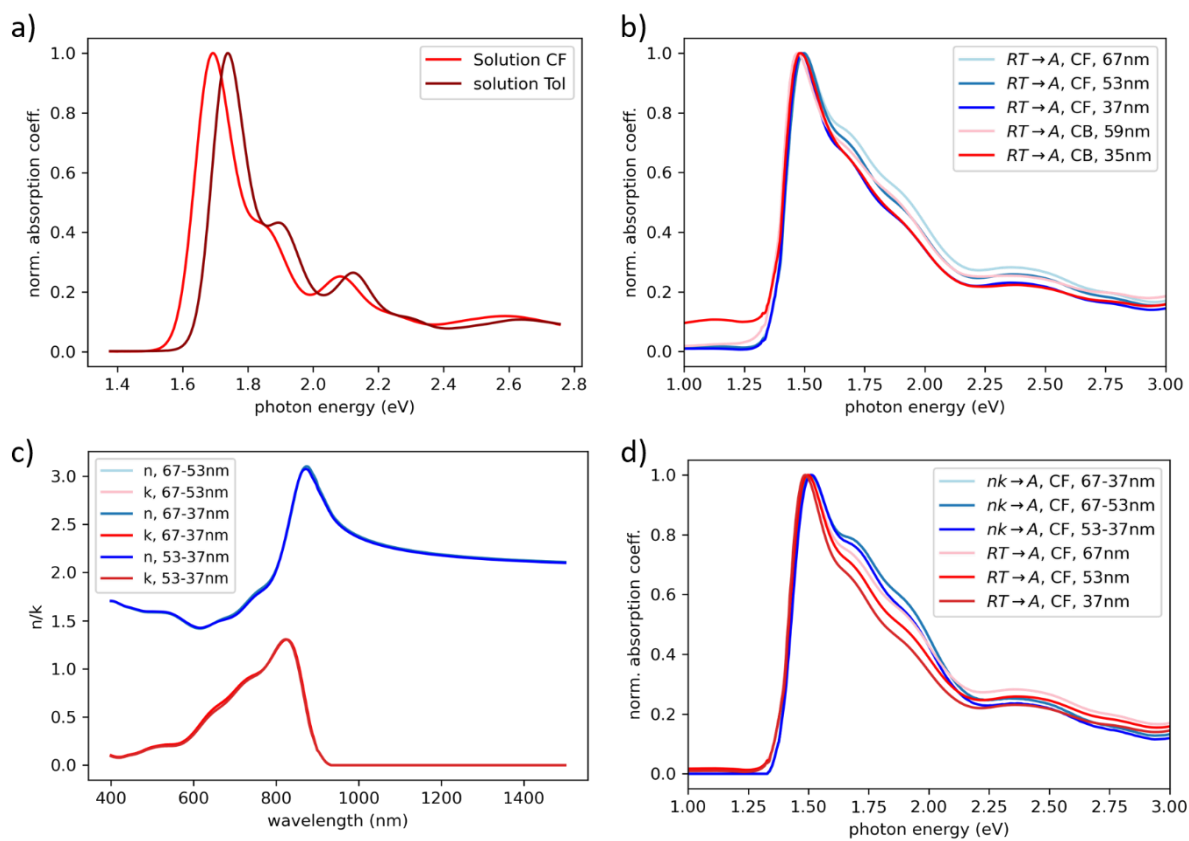

**Figure S11: (a)** Normalized Y6 solution absorption spectra prepared in CF and toluene. **(b)** Y6 normalized thin film absorption spectra determined using the Lambert Beer law prepared from CF (blue) and CB (red). **(c)** Real (blue) and complex (red) part of the complex index of refraction of the Y6 thin films determined using Kerremans *NK_Finder* program for the CF processed films. **(d)** Comparison between the normalized absorption spectra obtained from the Lambert Beer law and the attenuation coefficient for the CF processed films.

We note that CB processed Y6 film showed quite different structure order according to the literature and the film aggregates in domains adopting random orientations and lower crystallinity.[38,39] The absorption spectrum of CB-processed Y6 in Fig. S11b shows a slightly smaller intensity of the 0-1 band compared to the corresponding CF-processed sample and a slightly less broaden spectrum. We speculate that this might be a consequence of weaker FE-CT interactions and mixing in the more structurally disordered CB-processes sample.

### Section S14: Raman and ATR-FTIR.
**Raman Experimental**
Raman measurements are performed as explained in Methods. As an electronic transition is being excited, photoluminescence will also occur; this leads to a strong 'fluorescence background' being observed in the Raman spectrum. In addition, it is only the vibrational modes coupled to the electronic transition which are resonantly enhanced. This can simplify the spectrum, but can be misleading.

Because of the broad absorption of NFAs across the visible and NIR, it is not possible to perform 'true' non-resonant Raman spectroscopy with any of the available excitation lasers. However, by selecting a 532 nm excitation laser, the absorption and fluorescence background could be reduced (but not eliminated). In addition, the scattering intensity, which is proportional to $\lambda^{-4}$, is maximised with this choice of laser.

**ATR-FTIR Experimental**

In the experimental setup used here, the broadband IR beam interacts with the sample via attenuated total reflectance (ATR) as explained in Methods. The sample is in direct contact with a high refractive index crystal, and the IR beam undergoes several total internal reflections at the boundary between the crystal and the sample. An evanescent wave extends into the sample at each point of reflection, providing the opportunity for IR absorption. Unlike in transmission geometry, the FTIR spectrum can be acquired for thin films prepared on relatively IR-opaque substrates in ATR geometry, with no further preparation required. This crucially avoids the requirement to prepare additional thin films on expensive, IR-transparent calcium fluoride substrates, and allows for direct and valid comparison with the samples used in other spectroscopic techniques.

Due to the very low thickness of the film, there was significant penetration of the evanescent wave through the sample and into the substrate. This results in a broad, relatively featureless background absorption from the substrate, whose absorption increases with decreasing wavenumber. The sharp, narrow peaks of the sample IR absorption are easily distinguishable on top of this background. However, for the avoidance of doubt we have presented the background-corrected FTIR spectra only in the range $\tilde{\nu} \geq 1500 \text{cm}^{-1}$ where the substrate absorption is relatively low.

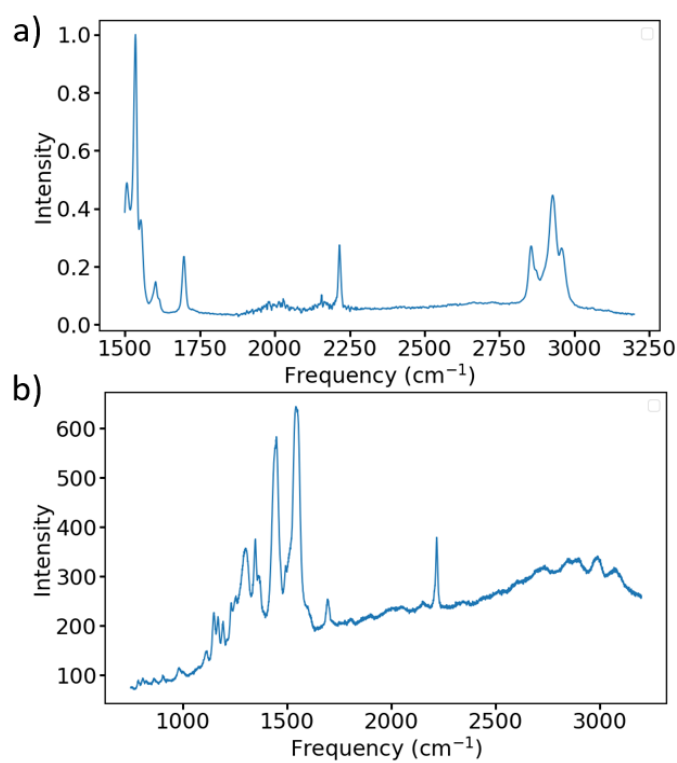

**Figure S12:** (a) ATR-FTIPR spectrum and (b) Raman spectrum.

## Section S15: Transient absorption spectroscopy of Y6 in dilute media.

**Y6 Toluene Visible Probe Picosecond Transient Absorption Spectroscopy**

Fig. S13 displays the picosecond transient absorption of a diluted solution of Y6 in toluene. Within the spectrum, three bands can be observed. The first is a central ground state bleach (GSB) ranging from 575 to 810 nm. Additionally, two photo-induced absorption (PIA) bands are present, with one centred at 530 nm (weak) and the other at 875 nm (strong). These PIAs are attributed to the $S_1$ state. The GSB exhibits a considerable amount of vibronic structure, similar to that found in the steady-state absorption spectrum shown in Fig. 2 of the main text. Alongside the overall decay of the GSB, there is a gradual shift towards shorter wavelengths, specifically at the blue edge, occurring over a period of approximately 5 ps.

Considering next the kinetics in Fig. S13b, we observe a surprising level of complexity. A source of this is likely to be the solvent-solute interactions of the Y6 molecule in solution. By analysing the fitting of the biphasic kinetics of the GSB in the 735–745 nm range and the PIA in the 850–900 nm range as shown in Fig. S13c we can identify two essential timescales within the system: $\tau_1$, approximately 7 ps, and $\tau_2$, approximately 400 ps. We note that the faster timescale closely matches that previously observed for other donor-acceptor-type organic chromophores in toluene solution.[41] Thus, we attribute $\tau_1$ to a solvent relaxation process. $\tau_2$, on the other hand, is assigned to the decay of the exciton to the ground state. By contrast, when dispersing Y6 at 2 wt% in the wide gap host mCBP (Fig. S14), we note that the fast component of the spectral evolution is supressed in the rigid solid-state environment. Consequently, the kinetics from all spectral regions decay largely in-step with a time constant of approximately 350 ps. This further supports the hypothesis that the ~7 ps timescales spectral evolution seen in toluene is due to solvent-solute relaxation effects.

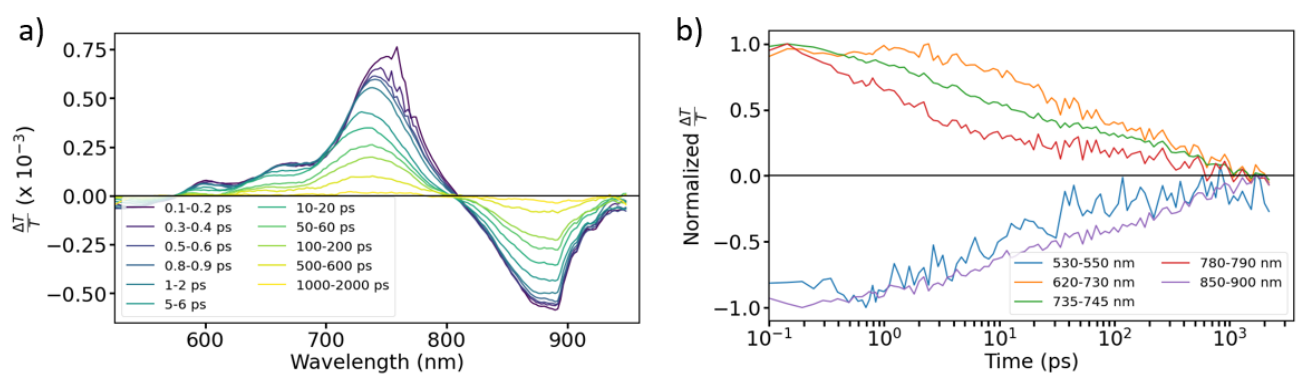

**Figure S13:** (a) Dilute Y6 toluene solution visible probe ps-TAS spectra, fluence = 4.97 µJ cm$^{-2}$ ($\lambda_{pump}$ = 765 nm). (b) Related kinetics normalized to maximum.

**Y6 2 wt% in mCBP Film Visible Probe Picosecond Transient Absorption Spectroscopy**

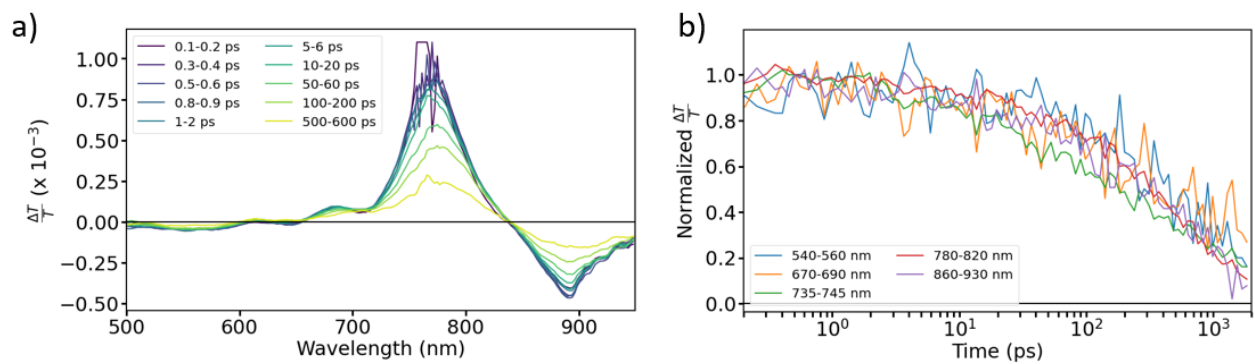

**Figure S14: (a)** Y6 2 wt% in mCBP film visible probe ps-TAS spectra, fluence = 7.6 µJ cm$^{-2}$. ($\lambda_{pump}$ = 765 nm). **(b)** Related kinetics normalized at 0.5 ps.

**Y6 50 wt% in mCBP Film Visible Probe Picosecond Transient Absorption Spectroscopy**

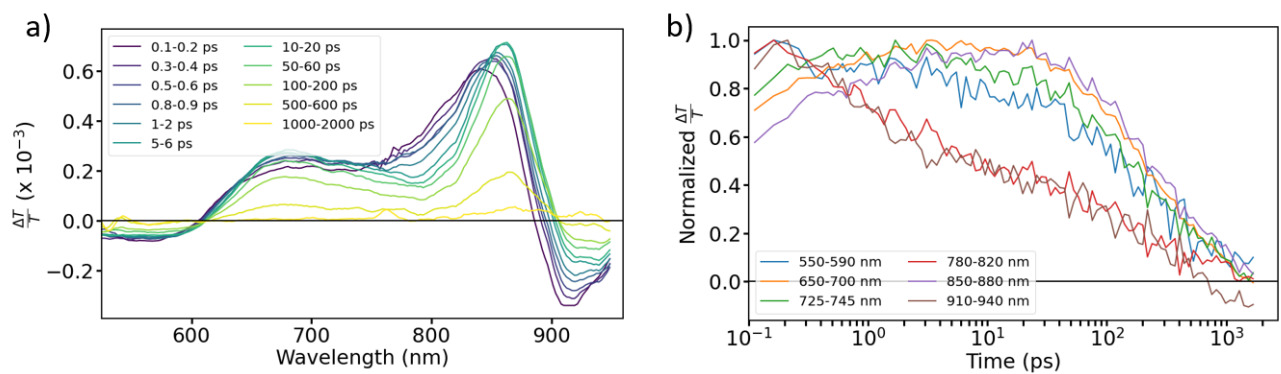

**Figure S15: (a)** Y6 50 wt% in mCBP film visible probe ps-TAS spectra, fluence = 1.0 µJ cm$^{-2}$. ($\lambda_{pump}$ = 765 nm). **(b)** Related kinetics normalized to maximum.

## Section S16: Transient absorption spectroscopy of Y6 in BHJ blends.

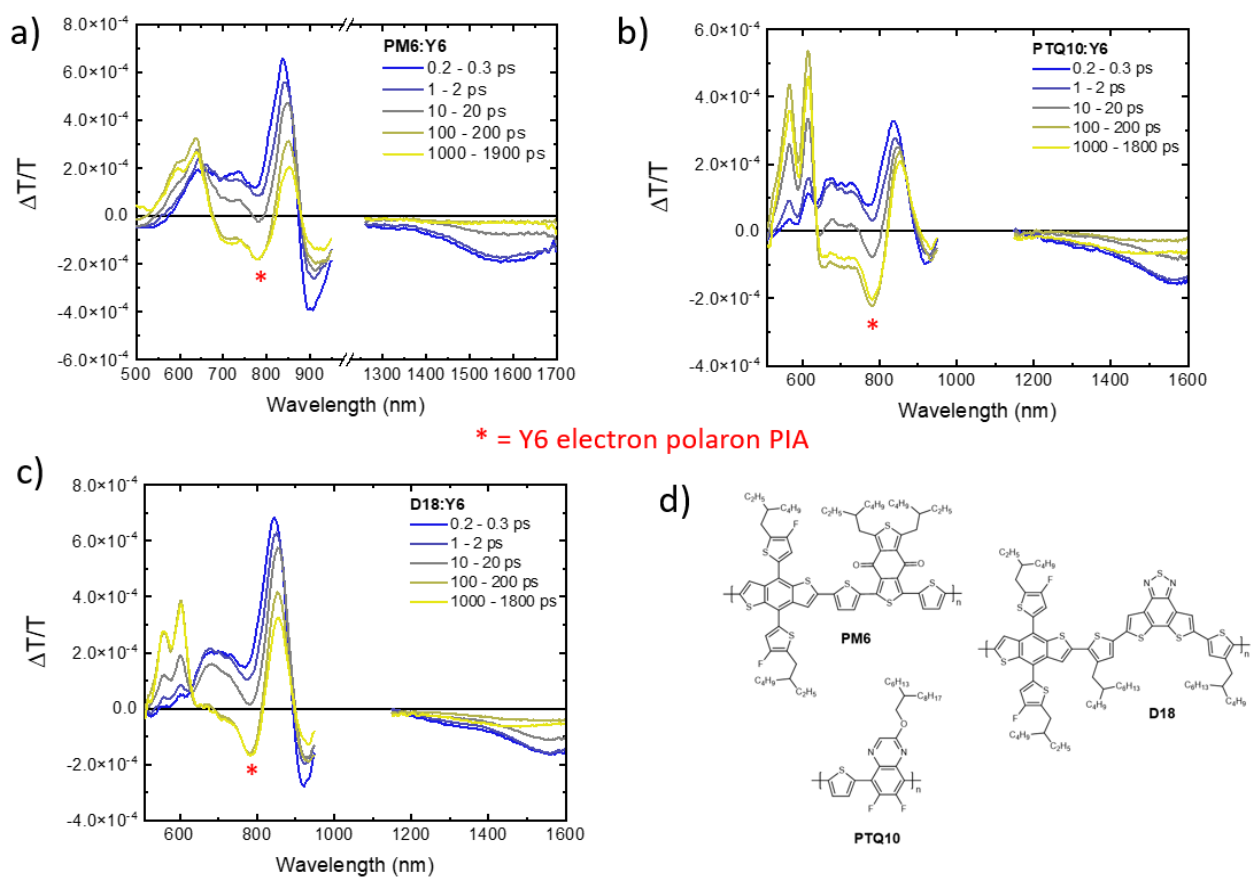

**Figure S16:** Transient absorption spectra of BHJ blends of Y6 with commonly used electron donating polymers: **(a)** PM6:Y6, **(b)** PTQ10:Y6 and **(c)** D18:Y6. ($\lambda_{pump}$ = 800 nm). We note that the hole polaron PIA of the donor polymers has previously been reported at 920 nm (PM6), 910 nm (PTQ10), and 890 nm (D18).[42] Thus, we can assign the new PIA formed at 780 nm over the timescales of charge transfer to the electron polaron on Y6. **(d)** Donor Polymers chemical structures.

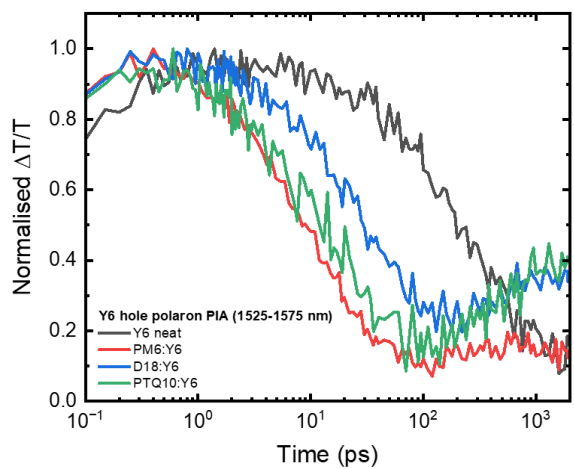

**Figure S17:** Kinetics of BHJ blends of Y6 with commonly used electron donating polymers. The 1550 nm PIA in Y6 (*i.e.* the hole polaron) is clearly quenched via hole transfer in the blends. The later rise in blends is due to formation of the overlapping Y6 triplet PIA via non-geminate recombination.[5]

# Section S17: Additional Transient Absorption spectroscopy data on Y6 films.

## Y6 Neat Film Visible Probe picosecond Transient Absorption Spectroscopy

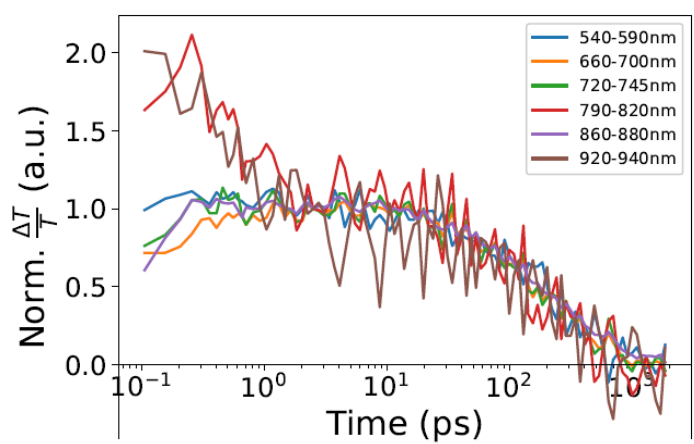

**Figure S18:** Ultrafast TA taken from Y6 singlet excitonic PIA at 920-940 nm. The kinetic falls to about ~40% of peak intensity by 2 ps – allows us to estimate what fraction of the photoexcitations form polaronic states vs remain as excitonic. Y6 neat film visible probe ps-TAS spectra, normalized at 2.5 ps (fluence = 0.61 µJ cm$^{-2}$, $\lambda_{pump}$ = 765 nm).

## Y6 Neat Film Visible Probe Femtosecond Transient Absorption Spectroscopy

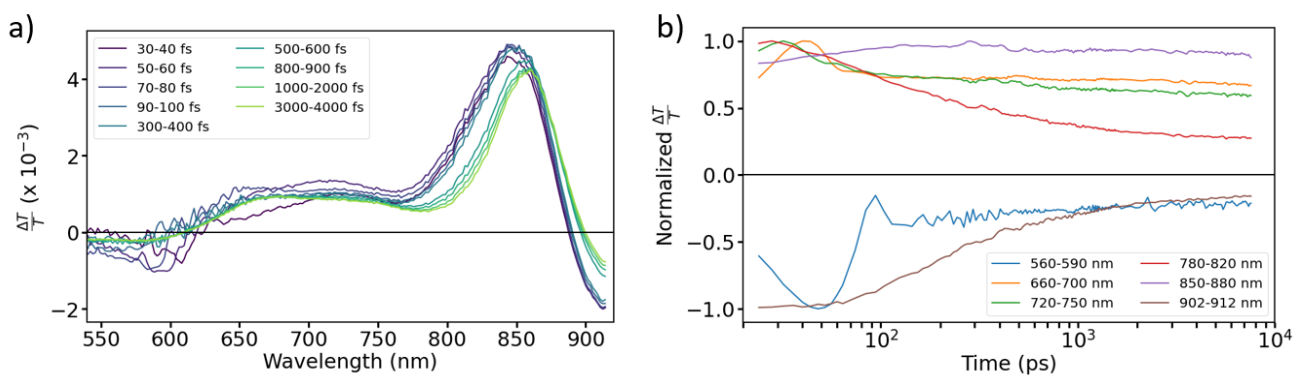

**Figure S19: (a)** Y6 neat film visible probe fs-TAS spectra. ($\lambda_{pump}$ = 580 nm). **(b)** Related kinetics normalized to maximum. Biexponential fits of selected Y6 neat film visible probe fs-TA kinetics at **(c)** 780-820 nm and **(d)** 902-912 nm.

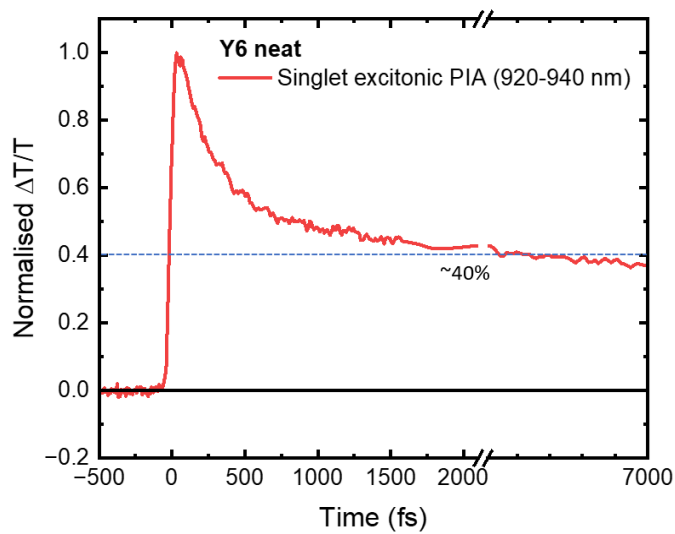

**Figure S20:** Ultrafast TA taken from Y6 singlet excitonic PIA at 920-940 nm. The kinetic falls to about ~40% of peak intensity by 2 ps – this allows us to estimate what fraction of the photoexcitations form polaronic states vs remain as excitonic.

**Y6 Neat Film Near and Shortwave Infrared Probe (SWIR) Transient Absorption Spectroscopy**

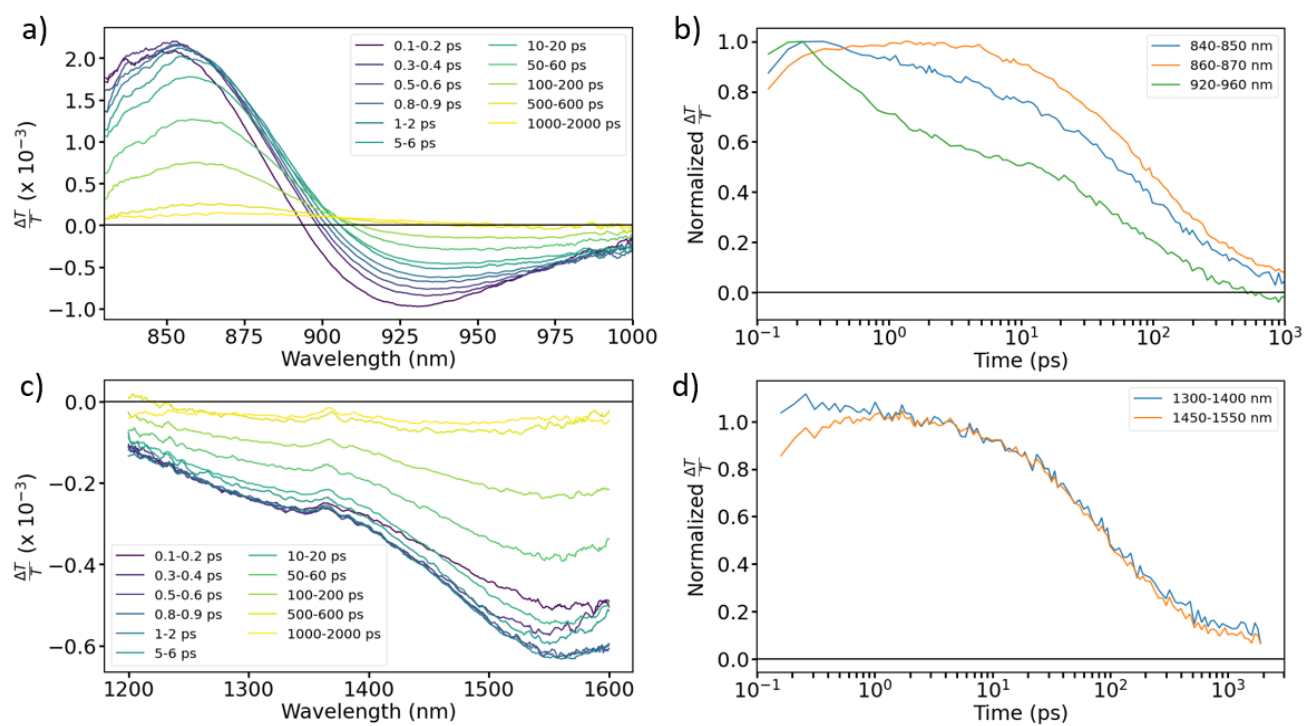

**Figure S21: (a)** Y6 Neat film NIR probe ps-TAS spectra, fluence = 1.11 µJ cm$^{-2}$. ($\lambda_{pump}$ = 800 nm). **(b)** Related kinetics normalized to maximum. **(c)** Y6 neat film SWIR probe ps-TAS, fluence = 2.3 µJ cm$^{-2}$ ($\lambda_{pump}$ = 800). **(d)** Related kinetics normalized to maximum.

**Y6 Neat Film Selective Excitation of Electronic States**

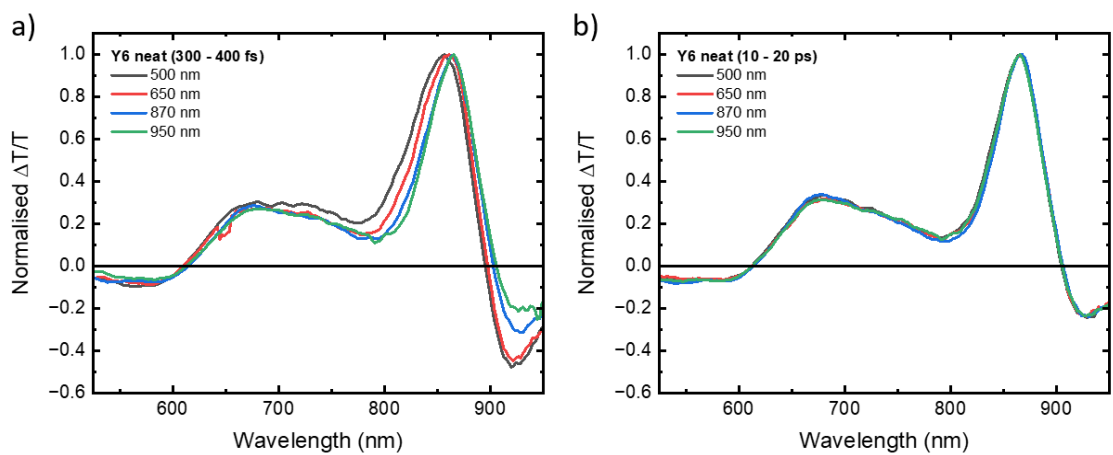

**Figure S22: (a)** Early time regime (0.3–0.4 ps) and **(b)** late time regime (10–20 ps) excitation-dependent Y6 neat film visible probe ps-TA spectra, normalised to the GSB maximum.

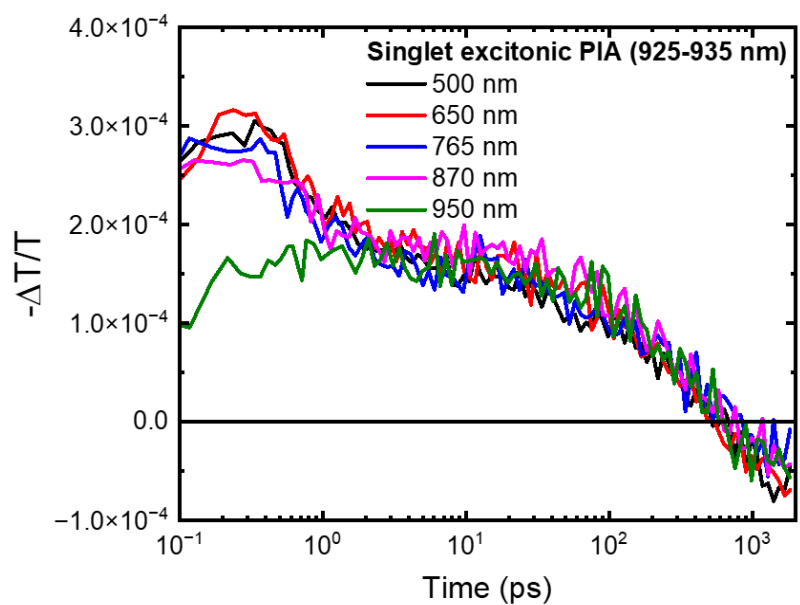

**Figure S23:** Kinetics taken from the singlet excitonic PIA (925 – 935 nm) for a range of excitation wavelengths.

## Section S18: Spectroelectrochemistry measurements

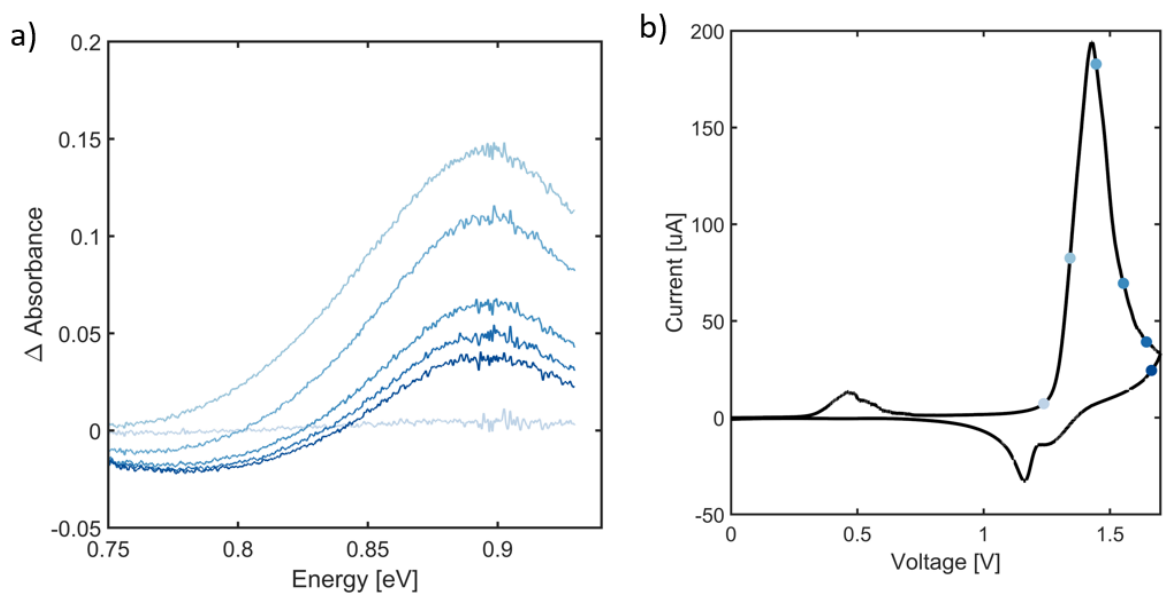

**Figure S24:** Spectroelectrochemistry on a neat Y6 film. The coloured dots on the cyclic voltammetry (CV) trace correspond to the spectra with the same colour.

## Section S19: Magnetic Resonance Spectroscopies

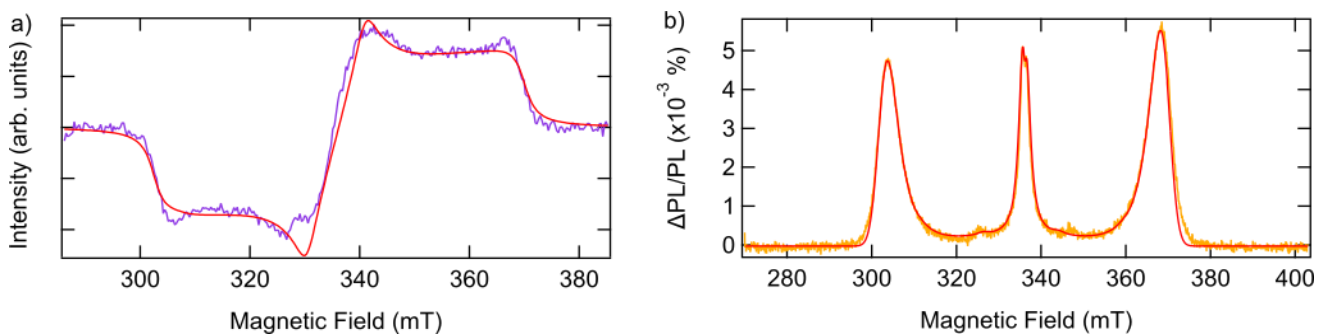

**Figure S25:** EasySpin simulations of all neat Y6 film from trEPR **(a)** and PLDMR **(b)** measurements, shown in the main text.

**Table S9.** Parameter for PLDMR spectral simulations using the MATLAB tool EasySpin for spin-coated substrates. Parameters given are the ZFS parameters $D$ and $E$, ordering factors $\lambda_\theta$, $\lambda_\phi$ and the spectral linewidth with Gaussian and Lorentzian contributions [Gaussian Lorentzian] and weight between for the relative signals. *: $E$ value cannot be determined due to high ordering.

| | Triplet | | | | | Polaron pair | |
|---|---|---|---|---|---|---|---|
| Material | $D/h$ [MHz] | $E/h$ [MHz] | $\lambda_\theta$, $\lambda_\phi$ | Lw [mT] | weight | Lw [mT] | weight |
| Y6 (PLDMR) | 950 | 150* 150 | 11.0, 0 0.0, 0 | 4.0, 0 2.0, 0 | 0.62 0.11 | 0.8, 2.5 0, 1.3 | 0.28 -0.05 |

**Table S10.** Parameter for trEPR spectral simulations using the MATLAB tool EasySpin for spin-coated substrates. Parameters given are the ZFS parameters *D* and *E*, relative zero-field populations, ordering factors λ_θ, λ_φ and the spectral linewidth with Gaussian and Lorentzian contributions [Gaussian Lorentzian].

| Material | $D/h$ [MHz] | $E/h$ [MHz] | $[p_z, p_y, p_x]$ | $\lambda_\theta, \lambda_\phi$ | Lw [mT] |
|---|---|---|---|---|---|
| Y6 | 945 | 215 | [0.00 0.66 0.34] | 0, 0 | [0 2] |